\pdfoutput=1 
\documentclass[%
aps,
superscriptaddress,
preprint,%
reprint,%
floatfix
]{revtex4-1}

\usepackage{amsmath}
\usepackage{comment}
\usepackage{dcolumn}

\usepackage{chngcntr}

\usepackage[colorlinks=true, allcolors=blue]{hyperref}

\usepackage{filecontents}

\usepackage{placeins}
\usepackage{color}
\usepackage{graphicx}
\usepackage{epstopdf}

\newcommand{\pten}[2]{${#1\times10^{#2}}$}
\newcommand{\con}[2]{\pten{#1}{#2}~cm$^{-2}$}
\newcommand{\vb}{${V_{\text{b}}}$}
\newcommand{\tc}{${T_{\text{c}}}$}

\newcommand{\muc}{${\mu^{*}_{\text{c}}}$}

\usepackage[caption=false]{subfig}

\begin{document}

\author{A. V. Lugovskoi}
\email{alugovskoi@science.ru.nl}
\affiliation{\mbox{Institute for Molecules and Materials, Radboud University, Heijendaalseweg 135, NL-6525 AJ Nijmegen, The Netherlands}}
\author{M. I. Katsnelson}
\affiliation{\mbox{Institute for Molecules and Materials, Radboud University, Heijendaalseweg 135, NL-6525 AJ Nijmegen, The Netherlands}}
\affiliation{\mbox{Theoretical Physics and Applied Mathematics Department,
Ural Federal University, 620002 Ekaterinburg, Russia}}
\author{A. N. Rudenko}
\affiliation{School of Physics and Technology, Wuhan University, Wuhan 430072, China}
\affiliation{\mbox{Institute for Molecules and Materials, Radboud University, Heijendaalseweg 135, NL-6525 AJ Nijmegen, The Netherlands}}
\affiliation{\mbox{Theoretical Physics and Applied Mathematics Department,
Ural Federal University, 620002 Ekaterinburg, Russia}}

\title{Electron-phonon properties, structural stability, and superconductivity of doped antimonene}

\begin{abstract} 
Antimonene is a recently discovered two-dimensional semiconductor with exceptional environmental stability,
high carrier mobility, 
and strong spin-orbit interactions. 
In combination with electric field, the latter provides an additional degree of control over the material's properties because of induced spin splitting. 
Here, we report on a computational study of electron-phonon coupling and superconductivity in $n$- and $p$-doped antimonene, where we pay a special attention on the effect of the perpendicular electric field. 
The range of accessible hole concentrations is significantly limited by the dynamical instability, associated with strong Fermi-surface nesting. At the same time, we find that in the case of electron-doping antimonene remains stable and can be turned into a state with strong electron-phonon coupling, with the mass enhancement factor $\lambda$ of up to 2.3 at realistic charge carrier concentrations.
In this regime, antimonene is expected to be a superconductor with the critical temperature of $\approx$16~K.
Application of bias voltage leads to a considerable modification of the electronic structure, affecting the electron-phonon coupling in antimonene.
While these effects are less obvious in the case of electron-doping, the field effect in hole-doped antimonene results in a considerable variation of the critical temperature, depending on bias voltage.

\end{abstract}

\maketitle

\section{Introduction}

Antimonene is a recent addition~\cite{Zhang2015,Lei2016} to a growing family of elemental two-dimensional (2D) materials. This monolayer phase of antimony adopts a buckled honeycomb lattice ($D_{3d}$ point group), similar to those of 
silicene~\cite{Vogt2012} and germanene~\cite{Zhang2016}, or more recently discovered blue phosphorene~\cite{Gu2017}.
Structurally, antimonene is identical to a single layer of bulk antimony, which possesses layered rhombohedral structure with a $D_{3d}$ point group.
Antimonene was successfully obtained by various experimental techniques~\cite{Ares2018}, including but not limited to, epitaxial growth~\cite{Wu2017}, liquid- and solid-phase exfoliation~\cite{Gibaja2016,Ares2016}.
Based on both experimental observation and \textit{ab initio} modeling~\cite{Ares2016,Zhang2015}, antimonene is believed to be a material with remarkable stability in air and in water~\cite{Ares2016,Wu2017}. 
This alone makes antimonene an appealing candidate for various applications because instability at realistic conditions is one of the main factors limiting the application of 2D materials~\cite{Morishita2015,Kuriakose2018}.
From a fundamental point of view, characteristic feature of antimonene is a strong spin-orbit coupling (SOC)~\cite{Rudenko2017}, with the intraatomic SOC constant $\lambda_{\text{SOC}}=0.34$ eV. A proper account of SOC is important for a correct description of the electronic structure and effective masses~\cite{Rudenko2017}. Antimonene is expected to have high charge carrier mobility~\cite{Pizzi2016} comparable to or exceeding that of the other 2D semiconductors. Besides, monolayer Sb is an indirect band semiconductor with a theoretically estimated band gap of 1.2 eV \cite{Singh2016}. 
These properties make antimonene suitable for application in electronic~\cite{Pizzi2016} and optical devices~\cite{Singh2016}, where thickness- and strain-tunable band gap could provide an additional control over the material's properties~\cite{Zhao2015,Zhang2015}.

Superconductivity is not uncommon for two-dimensional materials~\cite{Uchihashi2017}. It is observed in thin films~\cite{Ozer2006}, atomic sheets~\cite{Ludbrook2015}, surface atomic layers~\cite{Ge2015} and other 2D structures of various chemical compositions.
Recent experimental works~\cite{Chapman2016,Ludbrook2015} revealed superconductivity in graphene laminates with the critical temperatures \tc~in the range of 4--6~K.
Magnetization measurements of intercalated black phosphorus reveal \tc~$=3.8\pm$0.1~K, which was shown to be the same for different intercalants~\cite{Zhang2017}.
Also, a large number of experimental measurements of \tc~were recently performed for doped 2D transition metal dichalcogenides: WS$_{2}$ and NbSe$_{2}$ were demonstrated to be superconducting below 3~K \cite{Lu2018,Xi2015}, while MoS$_2$ was reported to have \tc~in the range of 7--11~K \cite{Lu2015,Ye1193}.
Superconductivity in two dimensions is especially interesting in view of the possibility of controlling the electronic structure of 2D materials by
strain, electric field, thickness, or substrate. Superconducting 2D materials also appear as an appealing testbed for studying interface phenomena and proximity effects. In particular, 2D materials can be implemented as a part of the Josephson junction~\cite{Heersche2007,Yabuki2016}. 

Computational studies predicting the superconductivity in monolayer graphene~\cite{Profeta2012,Margine2014} and phosphorene~\cite{Shao2014} were preceding the experimental observations~\cite{Ludbrook2015,Chapman2016,Zhang2017}.
Also, superconductivity at nonzero charge doping has already been predicted for recently proposed arsenene~\cite{Kong2018} (monolayer As), as well as for silicene \cite{Durajski2014}. Experimental data on superconductivity of these materials are not available yet.
While theoretical studies lack realistic accounts of the experimental setup, they allow study of the underlying mechanisms of superconductivity, for example, the key contributions to electron-phonon coupling or important changes of electronic structure due to charge carrier doping~\cite{Ge2013}.
Additionally, there is an opportunity to study modifications of the electronic structure and electron-phonon coupling properties including \tc~in the presence of strain~\cite{Shao2014} or other external factors.  Neither calculations nor measurements of \tc~are available in the literature for doped antimonene.

In this paper, we present a systematic study of the electron-phonon coupling and conventional superconductivity in $n$- and $p$-doped antimonene. To this end, we use a combination of Density Functional Theory (DFT)~\cite{Hohenberg1964,Kohn1965}  and Density Functional Perturbation Theory (DFPT)~\cite{Gonze1997} in conjunction with the formalism of Maximally Localized Wannier Functions (MLWF)~\cite{Giustino2007,Marzari2012}.
Besides charge doping, we also consider the role of an electric field applied in the direction perpendicular to the atomic layer, which allows us to modify the band structure.
Considering the interplay of these effects is interesting for a number of reasons.
For example, the perpendicular electric field allows one to control the band gap and effective masses in silicene and germanene~\cite{Ni2012,Acun2015}, as well is in few-layer phosphorene~\cite{Rudenko2015,Kim723,Liu2015}, where it is possible to transform material from normal insulator to topological insulator or metal.
Besides this, perpendicular electric field breaks inversion symmetry, which, in combination with strong SOC, induces spin-splitting of both valence and conduction bands~\cite{Kormanyos2013}.  
This case is observed, for example, in experimental work~\cite{Ye1193}, where setup combining liquid and solid gating is used to study superconductivity in MoS$_{2}$ flakes, and ionic liquid naturally creates the environment for emergence of perpendicular electric field.
Since SOC in antimonene is strong, this situation is particularly interesting to study.
For these reasons, we study the dependence of the electron-phonon coupling on bias voltage.

The rest of the paper is organized as follows. We first present the theory and computational methods used to calculate the electron-phonon coupling and superconducting critical temperatures, and also discuss the relevant approximations (Sec.~II). We then give a description of the calculated electronic structure and phonon dispersion of antimonene (Sec.~III~A). Main results are presented in Sec.~III~B, where we discuss the obtained dependencies of \tc~and electron-phonon coupling strength, as well as superconducting transition temperatures on charge carrier concentrations, with a detailed consideration of the most important cases. The effect of a perpendicular electric field (bias voltage) is analyzed and discussed in Sec.~III~C. 
In Sec.~IV, we summarize our results and conclude the paper.

\section{Theoretical background and computational details}

\subsection{Electron-phonon coupling and Allen-Dynes-McMillan equation}
In this section, we give a short theoretical description of the electron-phonon coupling and its relation to superconductivity.
More detailed description of the theory from the viewpoint of \textit{ab initio} calculations can be found in Ref.~\onlinecite{Giustino2017}.

The interaction of electrons with phonons is described by the matrix elements
\begin{equation}
g_{mn,\nu}({\bf k,q}) = \bigg( \frac{\hbar}{2m_0\omega_{{\bf q}\nu} }
\bigg)^{1/2} M_{mn}^{\nu}({\bf k},{\bf q}),
\label{eq:gmn}
\end{equation}
where 
\begin{equation}		
M_{mn}^{\nu}({\bf k},{\bf q}) =  
\langle \psi_{m{\bf k+q}} | \partial_{{\bf q}\nu}V | \psi_{n{\bf k}}\rangle,
\label{eq:mmn}
\end{equation}
and $\psi_{n{\bf k}}$ is the electronic Bloch function for band $n$ and wavevector ${\bf k}$; $\partial_{{\bf q}\nu}V$ is the derivative of the self-consistent potential associated with  the phonon of wavevector ${\bf q}$, branch index $\nu$; and frequency $\omega_{{\bf q}\nu}$; and $m_{0}$ is the atomic mass. 

It is useful to define a dimensionless representation of the electron-phonon coupling associated with the single phonon mode $\nu$ and wavevector ${\bf q}$ as an average over the Fermi surface:

\begin{equation}
\label{eq:lambda}
\begin{split}
\lambda_{{\bf q}\nu} = 
\frac{1}{N_{\rm F}\omega_{{\bf q}\nu}}\sum_{mn,{\bf k}} 
\mathrm{w}_{{\bf k}} |g_{mn,\nu}({\bf k,q})|^2 \\ 
\times\delta(\varepsilon_{n{\bf k}}-\varepsilon_F)\delta(\varepsilon_{m{\bf k}+{\bf q}}-\varepsilon_F),
\end{split}
\end{equation}							
which is closely related to the Eliashberg electron-phonon spectral function $\alpha^{2}F(\omega)$
\begin{equation}
\alpha^{2}F( \omega) = \frac{1}{2} \sum_{{\bf q}\nu} 
\mathrm{w}_{\bf q} \omega_{{\bf q}\nu} \lambda_{{\bf q}\nu}
\delta (\omega - \omega_{{\bf q}\nu}) 
, 
\label{eq:a2f}
\end{equation}											
where $\mathrm{w}_{\bf k}$ ($\mathrm{w}_{\bf q}$) is the symmetry-dependent weight of ${\bf k}$ (${\bf q}$) points, $\varepsilon_{n{\bf k}} (\varepsilon_{m{\bf k+q}})$ is the electronic energy, and $N_{\text{F}}$~ is the density of states (DOS) at the Fermi energy, $\varepsilon_F$.

Besides $\lambda_{{\bf q},\nu}$, we also consider the nesting function $\xi_{\bf q}$ as defined in~\cite{Bazhirov2010}:
\begin{equation}
\xi_{\bf q}= 
\sum_{mn,{\bf k}} 
\mathrm{w}_{{\bf k}}
\delta(\varepsilon_{n{\bf k}}-\varepsilon_F)\delta(\varepsilon_{m{\bf k}+{\bf q}}-\varepsilon_F).
\label{eq:nesting}
\end{equation}		
While the nesting function is independent of electron-phonon coupling matrix elements, it provides the understanding of effects of Fermi surface topology on effective electron-phonon coupling. It is important to mention, that the singularities of nesting function, associated with such topological phenomena as Fermi surface nesting may result in a high electron susceptibility at specific \textbf{q}, thus increasing the value of $\lambda$. Additional effects can be related to softening of phonon frequencies near this \textbf{q}-points, due to the giant Kohn anomaly~\cite{Katsnelson1994}.

The critical temperature of the superconducting transition according to the McMillan equation~\cite{McMillan1968}, modified by Allen and Dynes~\cite{Allen1975} can be estimated as:
\begin{equation}
T_{\text{c}} = 
\frac{\omega_{\text{log}}}{1.2}
\exp\left(
-\frac{1.04(1+\lambda)}
{\lambda-\mu_{\text{c}}^{*}(1+0.62\lambda)}
\right), 
\label{eq:ADMC}
\end{equation}	
which is based on the superconductivity theory of Migdal and Eliashberg~\cite{Migdal1958,Eliashberg1960}. Although derivation of Eq.~(\ref{eq:ADMC}) contains a number of approximations, it is a reasonable starting point for the estimation of~\tc.
In Eq.~(\ref{eq:ADMC}), \muc~is the Morel-Anderson effective Coulomb potential~\cite{Morel1962},

\begin{equation}
\omega_{\text{log}}=
\exp\left[
\frac{2}{\lambda}\int_{0}^{\omega_{\text{max}}}
d\omega \frac{\alpha^{2}F( \omega)}{\omega} \log \omega
\right]
\label{eq:wlog}
\end{equation}
is the logarithmically averaged phonon frequency,
\begin{equation}
\lambda = \sum_{{\bf q}\nu} \mathrm{w}_{{\bf q}} \lambda_{{\bf q}\nu}
\label{eq:lambtot}
\end{equation}
is the total Fermi energy-dependent electron-phonon coupling strength, which is also known as the mass-enhancement factor~\cite{Peters2018}. In the present paper, \muc~is considered as a phenomenological parameter with the typical values in the range of 0.1--0.2 \cite{McMillan1968,Allen1975,Giustino2017}.
Among 2D materials the Coulomb pseudopotential was estimated more rigorously for monolayer graphene in Ref.~\onlinecite{Margine2014} (\muc~$=0.16$) and Ref.~\onlinecite{Profeta2012} (\muc~$=0.1$),  {\color{black} bilayer graphene (\muc~$=0.155$) in Ref.~\onlinecite{Margine2016}, as well as NbS$_2$ (\muc~$=0.2$) in Ref.~\onlinecite{Heil2017}.} We note that an accurate estimate of \muc~requires an account of dielectric screening, and, therefore, would be strongly influenced by the underlying substrate.

\subsection{Calculation details}
\label{subsec:methods}

The initial electronic structure and crystal structure optimization calculations were performed at the DFT level as implemented in the plane-wave {\sc Quantum ESPRESSO} ({\sc QE}) code~\cite{Giannozzi2009,Giannozzi2017},
using fully relativistic norm conserving pseudopotentials.
Exchange and correlation were treated with the local density approximation (LDA).
The kinetic energy cutoff for plane waves was set to 90 Ry, the Brillouin zone was sampled with a (16$\times$16) Monkhorst-Pack \textbf{k}-point mesh~\cite{Monkhorst1976}, and the electronic state occupancies were treated as fixed.
The crystal structure was fully relaxed with a threshold of \pten{1}{-12}~eV for total energies and \pten{1}{-12}~eV/\AA~ for forces.
The vacuum thickness of 30 \AA~was used to avoid spurious interactions between the supercell periodic images in the direction perpendicular to the 2D plane.
The Brillouin zone for phonons within DFPT was sampled by a (16$\times$16) \textbf{q}-point mesh, and the self-consistency threshold of \pten{1}{-16}~eV was used.

Electronic structure, dynamical, matrices and electron-phonon matrix elements obtained from DFT and DFPT calculations were used as the initial data for Wannier interpolation within the MLWF formalism, as implemented in {\sc EPW}~\cite{Giustino2007,Ponce2016}.
The calculation of the electron-phonon-related properties was performed on dense grids of (432$\times$432) \textbf{k}- and (208$\times$208) \textbf{q}-points, which ensures the numerical convergence of the results presented in this work.

\subsection{Role of charge doping, out-of-plane acoustic phonons, and bias voltage}
\label{subsec:approx}

In this work, we simulate the charge carrier doping of antimonene using the rigid band shift approximation, that leaves the electronic structure and phonon dispersion unchanged.
Typically, charge doping in 2D materials mainly affects the out-of-plane mode and optical phonons~\cite{Margine2014,Kong2018,Shao2014}.
At the same time, out-of-plane acoustic mode (ZA) is not taken into account in our calculations. 
Although consideration of ZA mode may be important for fundamental understanding of the associated effects, it is of little practical interest for the superconductivity studies: Interaction of 2D materials with a substrate would suppress out-of-plane vibrations making the already small coupling of electrons with these modes negligible. Furthermore, a correct description of such effects is not trivial within the slab geometry.
An accurate description of electrons coupling with out-of-plane phonon modes is presented in Ref.~\onlinecite{Sohier2017} for doped graphene, where it is also shown, that the contribution from ZA mode is negligible. 
Optical phonons, as we show below, have a minor effect on the electron-phonon coupling and~\tc. Therefore, the rigid band shift approximation is justified for the purpose of our study.
The charge carrier concentration $\delta \rho$ is thus chosen in accordance with the ground state DOS and the Fermi energy, $\delta \rho(\varepsilon_F)=\int_0^{\varepsilon_F}d\varepsilon \, \rho(\varepsilon)$.

We consider both $n$- and $p$-doping cases. We limit ourselves to the concentrations less than~\con{1}{15}.
Although such charge carrier concentrations correspond to a heavy doping regime, they are not unrealistic. Electron concentrations of the order of~\con{1}{15} are achievable in thin metallic films by means of electrochemical techniques~\cite{Daghero2012}. 
Smaller electron and hole concentrations of the order of \con{1}{14} can be achieved by liquid gating~\cite{Efetov2010} and  solid-electrolyte gating~\cite{Xu2017}.
{\color{black}At the same time, the presence of the strong Fermi-surface nesting~\cite{Antropov1988,Vaks1989,Katsnelson1994}, as well as high charge carrier concentrations may lead to the loss of structural stability of the system.
Since such instabilities cannot be detected within the rigid band approximation, we also consider lattice dynamics of antimonene using the jellium doping method, as implemented in the DFT code used.
For every doping case, relaxation of atomic positions is performed, while the lattice parameter of undoped antimonene was used to model the behavior of the system on a substrate.}

To simulate the effect of a perpendicular electric field, we consider a tight-binding Hamiltonian obtained in the MLWF basis, and add position-dependent bias voltage, yielding the following Hamiltonian
\begin{equation}
H=\sum_{ij}t_{ij}c_i^{\dag}c_j+V_{\text{b}}/d\sum_{i}z_ic_i^{\dag}c_i,
\end{equation}
where the sum runs over the real-space MLWF orbitals $i$ and $j$, $c_i^{\dag}$,$c_i$ ($c_{j}$) are the creation and annihilation operators of electrons on the corresponding orbitals, $z_i$ is the $z$-component of the position operator of the orbital $i$, $t_{ij}$ is the corresponding hopping integral, $d$ is the buckling constant, and $V_{\text{b}}$ is bias voltage applied to the upper and lower planes of antimonene. 
It is worth mentioning that since the consideration of electric field is not performed in the self-consistent manner, as is done, for example, in~Ref.~\onlinecite{Li2018}, additional screening arising due to the charge carrier doping is not taken into account in our calculations; thus, the bias effect is quantitatively overestimated.

\begin{figure}[ht]
	\centering
	\includegraphics[]{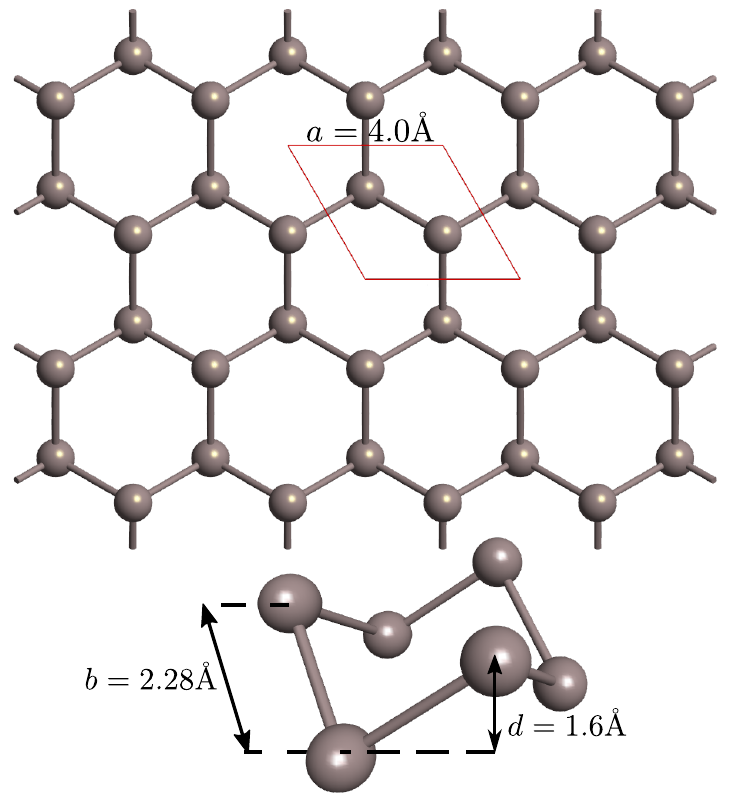} 
	\caption{Crystal structure of hexagonal antimonene; red lines denote the hexagonal unit cell, used in calculations. Lattice parameter, buckling constant, and bond length, denoted as $a$, $d$, and $b$ respectively, are also given on the figure.}
	\label{fig:structure}
\end{figure}
	
\section{Results and discussion}
\subsection{Electronic structure and phonon dispersion}
\label{subsec:dft}

\begin{figure}[ht]
	\centering
	\includegraphics[]{{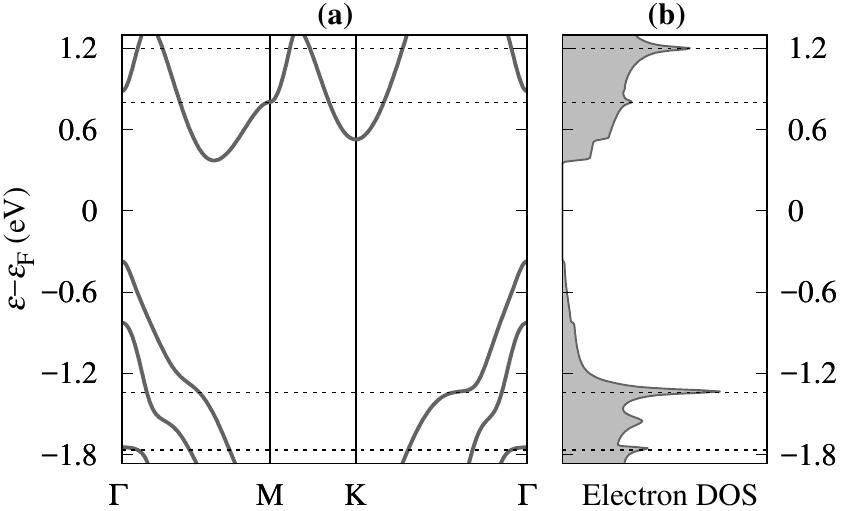}}\vspace*{0.3cm}
	\includegraphics[]{{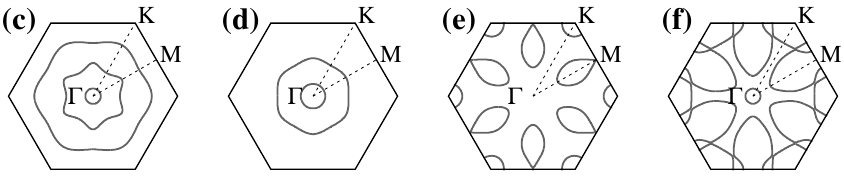}} 
    \caption{Electronic structure of antimonene. (a) Band structure plotted along high-symmetry directions of the Brillouin zone. (b) DOS in the relevant charge carrier concentration range. Horizontal lines on panels (a) and (b) mark the following charge carrier concentrations (from bottom to top): $N_{\text{h}}=$~\con{1}{15}, $N_{\text{h}}=$~\con{3.7}{14}, $N_{\text{e}}=$~\con{4}{14}, and $N_{\text{e}}=$~\con{1}{15}. The corresponding Fermi surface contours are given on subplots (c)--(f).}
	\label{fig:electrons}
\end{figure}

We find the calculated relaxed lattice parameter of free-standing antimonene to be $a = 4.0$~\AA~with the bond length $b = 2.82$~\AA~and buckling constant $d = 1.6$~\AA.
The structure and calculated parameters are illustrated in Fig.~\ref{fig:structure}.
The found values are consistent with previously reported data~\cite{Wang2015}.
The corresponding electronic band structure is given in Fig.~\ref{fig:electrons}(a), from which one can see that antimonene is an indirect gap semiconductor with the gap width of 0.7~eV. The obtained value is less than 1.0 and 1.2~eV obtained in Refs.~\onlinecite{Rudenko2017,Singh2016}, due to the differences in exchange-correlation functional, but agrees well with the value reported in~Ref.~\onlinecite{Wang2015}. 
The band structure and DOS [Fig.~\ref{fig:electrons}(b)] display high degrees of electron-hole asymmetry even at small Fermi energies, unlike those of typical 2D materials including graphene and phosphorene.

\begin{figure}[h]
	\centering
	\includegraphics[]{{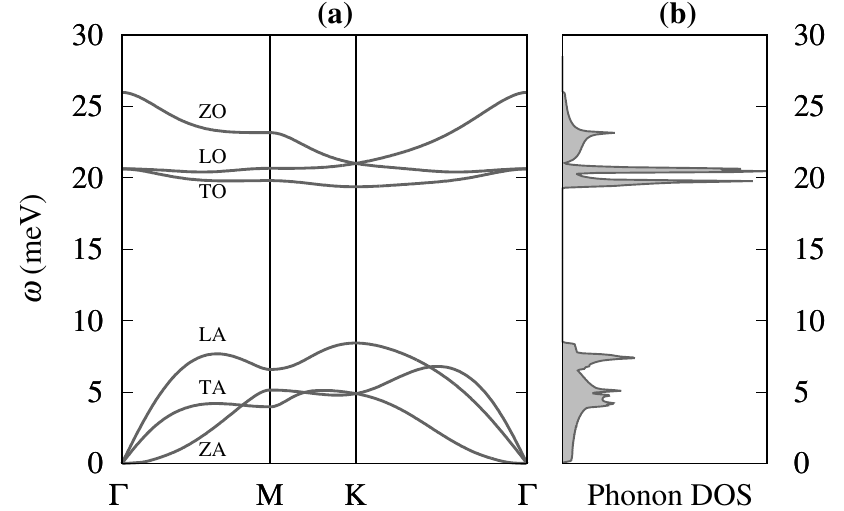}}
	\caption{Phonon specturm of antimonene. (a) Phonon dispersion plotted along high-symmetry directions of the Brillouin zone. (b) Phonon DOS.}
	\label{fig:phonons}
\end{figure}

At negative Fermi energies (hole-doping) the Fermi surface is initially formed by one valence band. As the Fermi energy reduces, second and third occupied bands get involved, forming three concentric pockets [see Fig.~\ref{fig:electrons}(c)].
In the \textit{K}--$\Gamma$ direction of the electronic structure one can see a flat region in the valence band, which gives rise to a van Hove singularity (VHS) in DOS ($\varepsilon=-1.3$~eV in Fig.~\ref{fig:electrons}). The corresponding
constant-energy contour is shown in Fig.~\ref{fig:electrons}(d), which exhibits a hexagonal warping.
This topology opens up a possibility for the scattering between the parallel regions of the surface, known as the Fermi surface nesting~\cite{Katsnelson1994,Landa2018}. In this case one can expect an increase of the electron-phonon coupling strength at the corresponding charge carrier concentrations ($N_{\mathrm h}=~$\con{3.7}{14}).
Further reduction of the Fermi energy leads to a widening of the contours 
accompanied by the bending of the hexagons, suppressing the nesting effect. 
The topology of the Fermi surface at positive Fermi energies (electron-doping) is determined entirely by a single conduction band, yet involving multiple valleys.
At small energies the surface is formed by six closed droplet-shaped pockets, originating from the valleys centered along the $\Gamma$--\textit{M} direction.
Additional circle-shaped pockets appear as the Fermi energy reaches the valley at the \textit{K} point [Fig.~\ref{fig:electrons}(e)].
Further increase of the Fermi energy results in a VHS originating from the band bending around the \textit{M} point.

At even higher conduction band fillings there is a distinctive change in the Fermi surface topology. Namely, previously closed pockets become connected, and an additional valley emerges around $\Gamma$ [circle region in Fig.~\ref{fig:electrons}(f)]. Similar to the valence band, one can see a hexagonal warping around the \textit{K} point, which gives rise to a VHS ($\varepsilon=1.2$ eV in Fig.~\ref{fig:electrons}), corresponding to a heavy electron-doping with $N_{\mathrm e}=~$\con{1.0}{15}.

Calculated phonon spectra is given on Fig.~\ref{fig:phonons}.
While the impact of SOC on the electronic structure is notable, it is significantly less prominent in the context of lattice dynamics.
In the long-wavelength limit, the frequency changes are nearly unnoticeable, introducing the difference below~1\%. At higher $\mathbf{q}$, particularly in vicinity of \textit{K} high-symmetry point, the difference in acoustic phonon frequency reaches~5\%.
Frequencies of optical phonon modes, particularly ZO and LO, demonstrate comparable changes.
Overall phonon spectra with SOC taken into account only slightly differs from that of without SOC (see the SM, Fig.~S1).
In comparison to other elemental 2D materials, antimonene has considerably lower phonon frequencies, which results, for example, in low thermal conductivity, as seen in~\cite{SWang2016}.
Elastic constants can be estimated from the frequencies of long-wavelength phonons, using the expression $\omega_\nu=q\sqrt{C_{ij}/\rho_{2D}}$, where $C_{ij}$ is the elastic constant, related to acoustic phonon mode $\nu$, $\rho_{2D}$ is the mass density of 2D material.
Out-of-plane phonon mode ZA with quadratic dispersion at low ${\mathbf q}$ is related to bending rigidity $\kappa$ in the following way $\omega_{ZA}=q^{2}\sqrt{\kappa/\rho_{2D}}$.
Thus, independent 2D elastic constants of antimonene would have the following values: $C_{11}=2.1$~eV/\AA$^{2}$, associated with LA phonon mode, and $C_{66}=0.8$~eV/\AA$^{2}$, associated with TA phonon mode, while $C_{12}=C_{11}-2C_{66}=0.5$~eV/\AA$^{2}$ and Young modulus $E = (C_{11}^{2}-C_{12}^2)/C_{11} = 2.17~\text{eV/\AA}^2$. The corresponding sound velocities are $v_{\text{s},LA}=3.4$~km/s and $v_{\text{s},TA}=2.1$~km/s. The bending rigidity associated with ZA phonons is $\kappa=0.3$~eV.
These values are significantly smaller than the elastic moduli and speed of sound of graphene, as well as black and blue phosphorene~\cite{DLiu2016,Zhu2014}.
Indeed, in contrast to light elements, heavy antimony atoms suppress vibrations, leading to smaller frequencies.
Finally, it is interesting to make a note of the role of anharmonic effects in antimonene. The characteristic cutoff wavevector below which anharmonic corrections become dominant is given by $q^{*}= \sqrt{
\frac{3T_{\text{R}}E}{16\pi\cdot\kappa^2}
}$.
At room temperature $T_{\text{R}}=300$~K it can be estimated as $q^{*}=0.2~\text{\AA}^{-1}$, which is an order of magnitude larger than for black phosphorus~\cite{Rudenko2016} and comparable with graphene~\cite{Katsnelson2010}.
In the context of superconductivity, however, these effects are not relevant.

\begin{figure*}[t]
	\centering
	\includegraphics[width=\textwidth]{{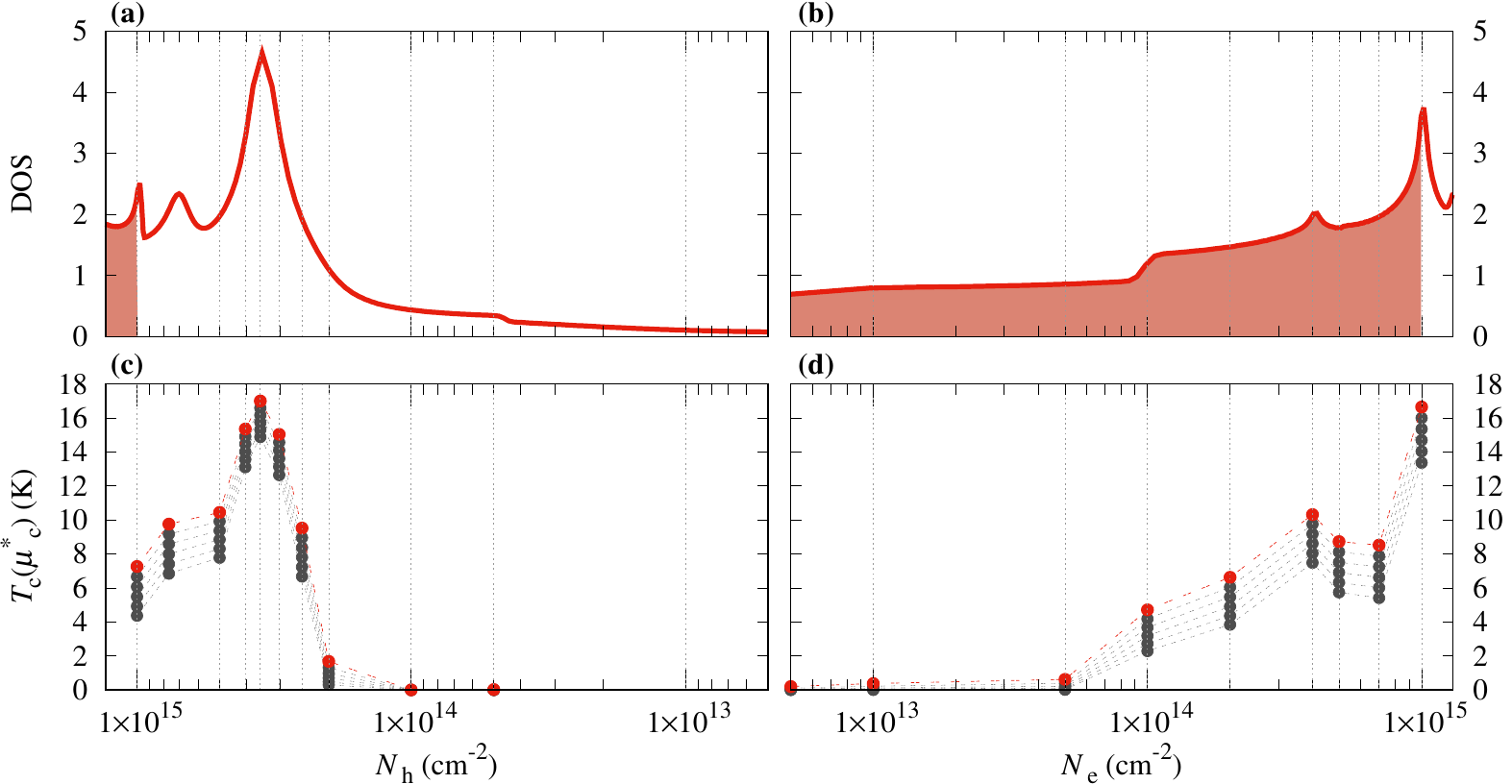}} 
	\caption{DOS [(a), (b)] and critical temperature [(c), (d)]  dependencies on charge carrier concentration for antimonene. Vertical dotted lines mark considered charge carrier concentrations. Multiple points for single $N_{\text{h/e}}$ on (c) and (d) represent \tc~at different values of \muc~(0.1, 0.12,...0.2) from top to bottom. Red symbols and lines on panels (c) and (d) mark \tc~values at \muc~$=0.1$. Dashed lines on bottom panels serve as a guide to the eye.}
	\label{fig:dos_tc_bias0}
    \vspace*{-0.3cm} 
\end{figure*}	

\subsection{Superconductivity and electron-phonon coupling}\label{subsec:main_res}

Calculated values of \tc~at various charge carrier concentrations are presented in Fig.~\ref{fig:dos_tc_bias0}. 
As can be seen, the concentrations resulting in the highest critical temperatures for both holes and electrons correlate with VHS of the corresponding charge carrier DOS. The highest value \tc~$=17$ K is achieved for the hole-doping with $N_\mathrm{h}=$\con{3.7}{14}.
In the case of electron-doped antimonene, the highest value is comparable (\tc~$\approx16$~K) yet it is observed at significantly higher charge carrier concentration $N_{\text{e}}=$~\con{1}{15}. In all considered cases, \tc~decreases by 2--3 K at the maximum chosen \muc~$=0.2$.

For the sake of comparison with experimental results, let us consider reference experimental concentrations reported recently.
In the context of superconductivity of intercalated graphene laminates, chemical doping was utilized leading to \tc~in the range 6--6.4~K at $N_{e}\approx$~\con{1}{14}~\cite{Ludbrook2015,Chapman2016}.
At this concentration antimonene demonstrates slightly lower \tc~right above the liquid helium temperatures (4.2~K). 
However, charge carrier concentrations achievable by the electrostatic doping~\cite{Zhao2017,Efetov2010,Xu2017}   yield higher critical temperatures: 6.6--10.4~K at 2--\pten{5}{14}~cm$^{-2}$ with a local maximum at~\con{4.1}{14}.
In case of the hole-doping, \tc~vanishes rapidly with increasing \muc~at charge carrier concentrations below \con{2}{14}, mainly due to small DOS.
In the vicinity of VHS, \tc~increases rapidly: At \con{3}{14} we find \tc~$=15.4$ K. 
Our estimation for \tc~in antimonene is comparable with that of phosphorene, according to the calculations reported in Refs.~\onlinecite{Shao2014,Ge2015}.
At $N_{\text e}=$~\con{7}{14} both materials yield \tc~$\approx10~$ K with phosphorene showing slightly higher values.
At smaller concentrations antimonene shows better results: At 1--\con{4}{14} electron-doped phosphorene is predicted to have \tc~in the range of 0.5--5~K, while antimonene exhibits \tc~$=10$~K.
Application of strain to phosphorene can change the situation: \tc~of phosphorene in this case can be tuned to exceed those of monolayer Sb in the aforementioned concentration range \cite{Shao2014,Ge2015}.
Calculated values of \tc~for the other group V 2D material, namely arsenene, closely resemble those of antimonene, with a slight shift to lower concentrations~\cite{Kong2018}.
Particularly, at concentrations ranging from \pten{0.9}{14} to \con{3.7}{14} $n$-doped arsenene shows \tc~in the range of 4.7--10.1~K, with the maximum value at \con{2.8}{14}, which is just 1~K higher, than for antimonene at the same concentration.
As for phosphorene, a strain tuning of \tc~is proposed for arsenene, which can increase these values.
\begin{figure*}[ht]
	\centering
	\includegraphics[width=\textwidth]{{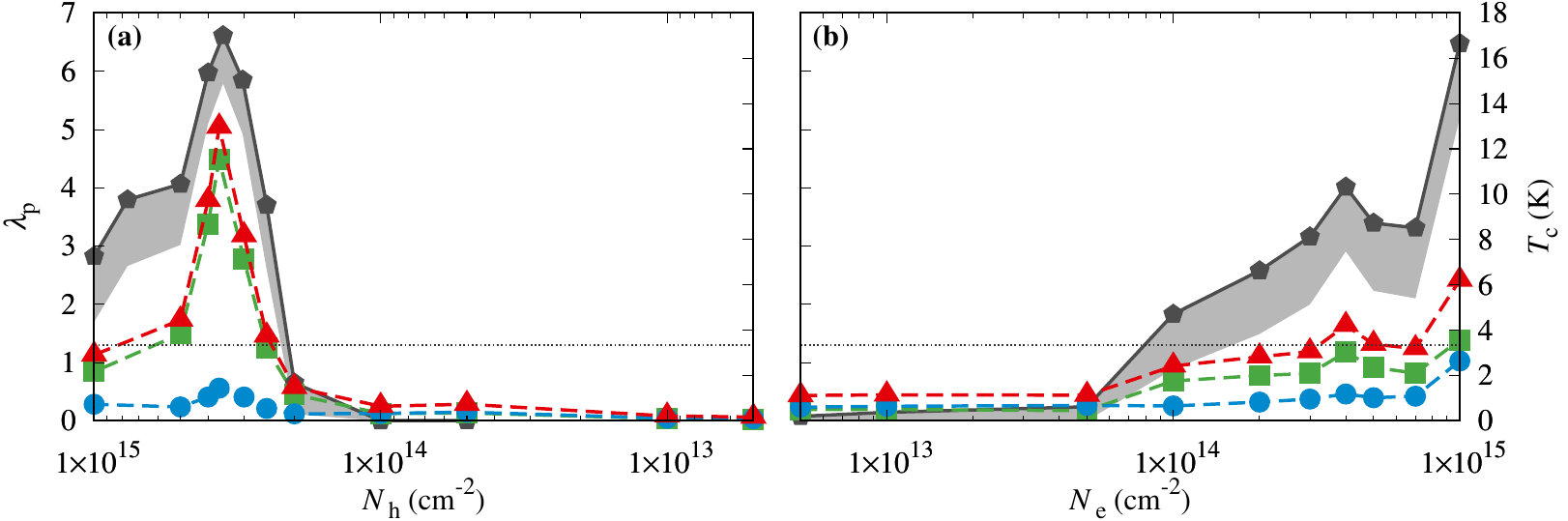}}
	\caption{Contributions to total electron-phonon coupling ($\lambda_\mathrm{p}$) and critical temperature ($T_{\mathrm{c}}$) as functions of hole (a) and electron (b) concentrations.
		Gray pentagons denote critical temperatures at $\mu^{*}$ = 0.1. Gray shaded region represents variability of \tc~with respect to $\mu^{*}$. Red triangles denote total electron phonon-coupling. Contributions to $\lambda_{\mathrm{p}}$ from acoustic and optical phonon modes are shown by green squares and blue circles, respectively.
		Dashed horizontal line corresponds to $\lambda=1.3$.
        }
	\label{fig:lambda_contrib_nza}
\end{figure*}
\begin{figure*}[ht]
	\centering
	\includegraphics[width=\textwidth]
	{{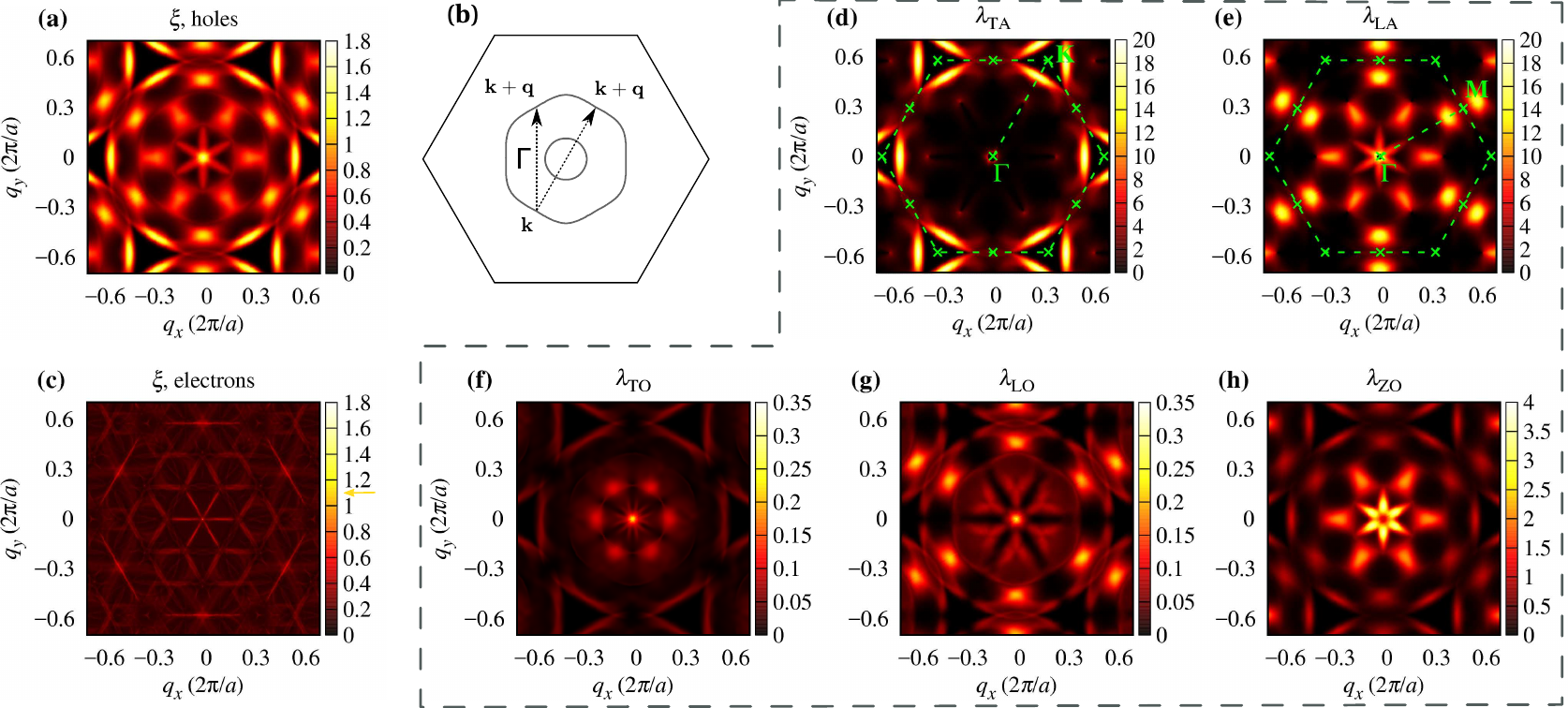}}
	\caption{Electron-phonon coupling and nesting function of doped antimonene. Panels (a) and (c) show nesting function resolved in $\bf q$-space for $N_{\mathrm{h}}=$~\con{3.7}{14} and $N_{\mathrm{e}}=$~\con{1}{15}, respectively. Heatmaps are given on the same scale for clarity, whereas actual maximum of $\xi$ is marked by a yellow arrow in panel (c). Panel (b) shows the Fermi surface contour for $N_{\mathrm{h}}=$~\con{3.7}{14}, where arrows mark two distinct electron scattering channels discussed in the text. In dashed frame, panels (d)--(g) show electron-phonon coupling contributions from different phonon modes resolved in $\bf q$-space. Green labels in panels (c) and (d) mark the Brillouin zone and high-symmetry directions. 
    }
	\label{fig:lambda_single}
    \vspace*{-0.3cm} 
\end{figure*}

Let us analyze electron-phonon coupling in antimonene and its role in the superconducting properties in more details.
Data on the averaged and doping-dependent electron-phonon coupling $\lambda$ [Eq.~(\ref{eq:lambtot})] of antimonene is given in Fig.~\ref{fig:lambda_contrib_nza} (red triangles).
It is worth mentioning that electron-phonon coupling strength of doped-antimonene is in general higher than that for other elemental monolayer materials discussed above. At comparable charge carrier concentrations, $\lambda$ in the range of 0.5--1.4 was obtained for phosphorene~\cite{Shao2014}, in the range of 0.76--1.27 for arsenene~\cite{Kong2018}, while $\lambda=0.6$ was measured experimentally for Li-doped graphene at $N_{\text{e}}=$~\con{1}{14}~\cite{Ludbrook2015}.
Electron-phonon coupling strength of antimonene exceeds these values already at $N_{\text{e}}=$~\con{1}{14}.
Despite significantly higher values of $\lambda$, this does not lead to a significant increase of \tc~in comparison with the other materials discussed.
It can be explained by the difference in characteristic phonon frequencies for these systems (see Sec.~\ref{subsec:dft}), and particularly smaller values of $\omega_{\text{log}}$ [Eq.~(\ref{eq:wlog})].
As a result, in the context of superconductivity this effect compensates for higher $\lambda$.
{\color{black}
The maximum value $\lambda=5$ is observed at $N_{\text{h}}=$~\con{3.7}{14}.
Thus, the case of strong electronic nesting is observed for the hole-doped antimonene, which leads to a strong electron-phonon coupling for certain wave vectors, and to a higher $T_{\text{c}}$.
At the same time, it has to be taken into account that the Fermi surface nesting, as well as Van Hove singularities in the electron energy spectrum leads to a general destabilization of the system~\cite{Antropov1988,Vaks1989,Katsnelson1994}.
The most obvious manifestation of this effect is the loss of structural stability. 
We discuss this aspect in Sec.~\ref{subsec:stability} in details.
}

Electron-doped antimonene demonstrates considerably smaller $\lambda$, yet larger than in the other elemental 2D materials: At $N_{\text{e}}>$~\con{3}{14} one has $\lambda>1.3$, particularly $N_{\text{e}}$=\con{1}{15} and ~\con{4}{14} give $\lambda=1.8$ and $\lambda=2.3$ respectively.
We note that in the original work where Eq.~(\ref{eq:ADMC}) was derived~\cite{Allen1975}, as well as in later works~\cite{Kresin1987}, it is pointed out that Eq.~(\ref{eq:ADMC}) underestimates \tc~for materials with strong electron-phonon coupling, characterized by $\lambda>1.3$.
Therefore, the critical temperatures reported here should be considered as a lower limit.

As can be seen from Fig.~\ref{fig:lambda_contrib_nza}, coupling with acoustic in-plane phonons (green squares) is the dominant contribution to $\lambda$. 
The coupling with optical phonons (blue circles) is small, but not negligible, at least in case of $p$-doping.
We now consider $\lambda_{\text{p}}$, contributions from individual phonon modes to the electron-phonon coupling at $N_{\text{h}}=$~\con{3.7}{14}, which are presented in Fig.~\ref{fig:lambda_single}.
The main contribution is provided by longitudinal acoustic mode, while coupling with TA mode gives the second highest value.
As it was mentioned before, at $N_{\text{h}}=$~\con{3.7}{14} there is an indication of a strong Fermi surface nesting. 
To further investigate these phenomena, we calculate the nesting function $\xi$ [Eq.~(\ref{eq:nesting})] shown in Fig.~\ref{fig:lambda_single}(a).
Indeed, $\xi$ exhibits maxima in the $\Gamma$--\textit{K} direction of ${\bf q}$, corresponding to the momentum transfer between parallel sections of the Fermi surface [dotted line in Fig.~\ref{fig:lambda_single}(b)]. As can be seen from Fig.~\ref{fig:lambda_single}(d), this mechanism provides a dominant contribution to the coupling with TA phonons. At the same time, one can also see another set of peaks in $\xi_{\bf q}$ along the $\Gamma$--\textit{M} direction. These peaks correspond to the momentum transfer between nonparallel parts of the Fermi surface [dashed line in Fig.~\ref{fig:lambda_single}(b)]. This mechanism turns out to be more important for the coupling with LA phonons,
which is shown in Fig.~\ref{fig:lambda_single}(e), resulting in a higher value of $\lambda_{\text{LA}}$.
The electron-phonon coupling strengths for optical modes are two orders of magnitude smaller than those for the acoustic ones, with the exception of optical out-of-plane ZO mode [Figs.~\ref{fig:lambda_single}(f)-\ref{fig:lambda_single}(h)].

\begin{figure*}[ht]
	\centering
	\begin{minipage}[c]{8.6cm}
		\centering
		\includegraphics{{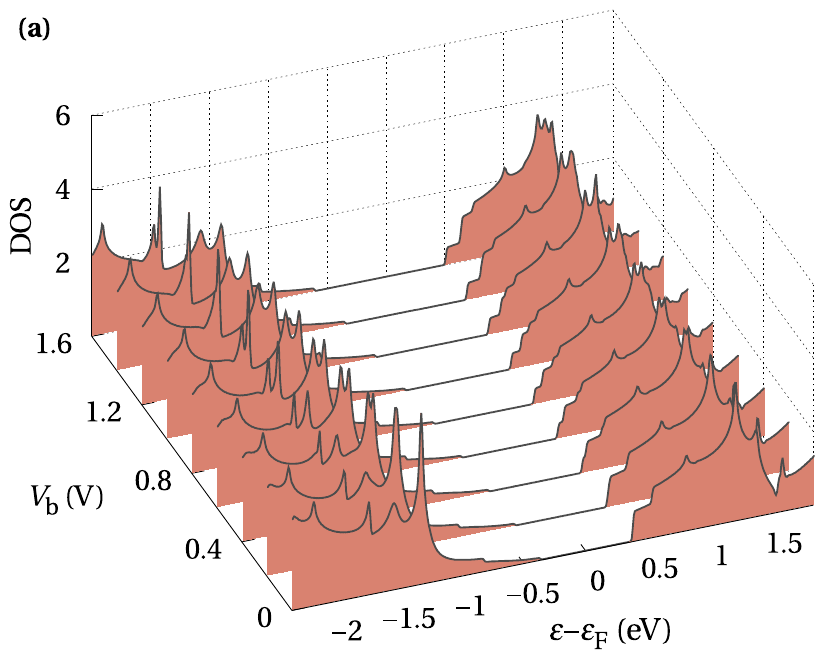}}
	\end{minipage}
	\begin{minipage}[c]{6.7cm}
		\centering
		\includegraphics{{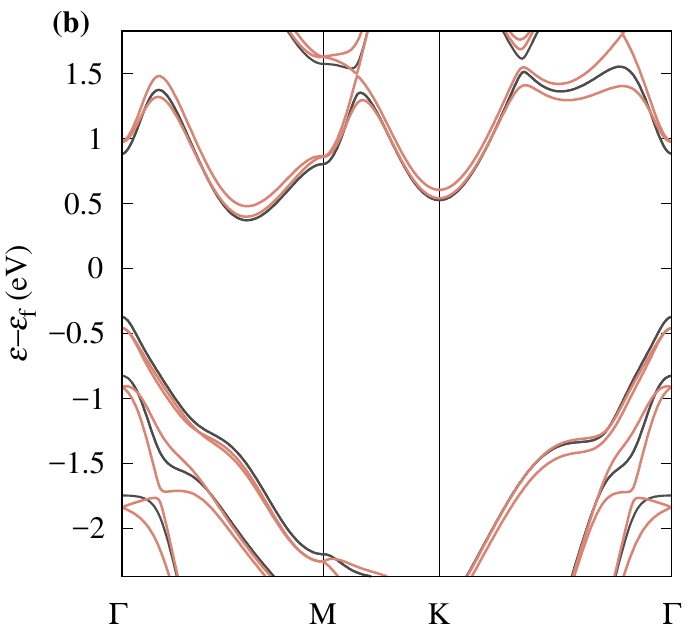}}
	\end{minipage}
	\caption{Effect of applied bias voltage on the electronic structure of antimonene. 
		(a) DOS at various values of bias voltage. 
		(b) Comparison of the electronic band structure for \vb=1.0~V (red line) and \vb=0.0~V (gray line).
	}
	\label{fig:dos_bias}
\end{figure*}

\begin{figure*}[ht]
	\includegraphics[width=0.9\textwidth]{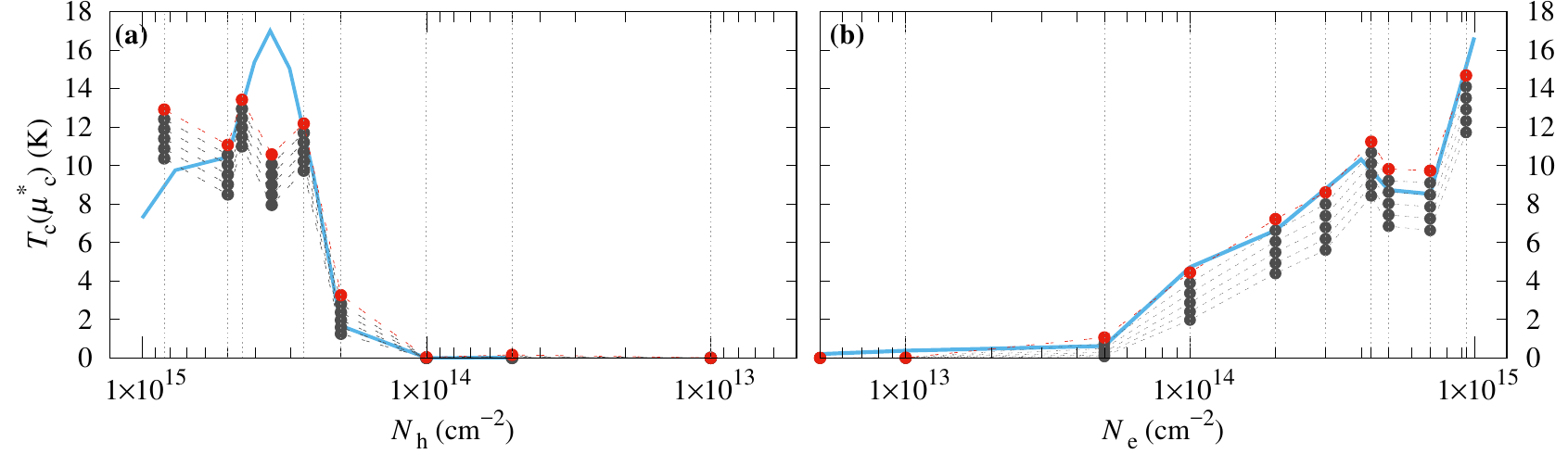}\vspace*{-0.2cm}
	\caption{Critical temperature dependence on charge carrier concentration at bias voltage of 1~V. Vertical lines mark relevant hole (a) and electron (b) concentrations. Six points for a single $N_{\text{h/e}}$ represent different values of \muc~(0.1, 0.12,...0.2). Blue solid line represents \tc~at \muc~$=0.1$ and \vb~$=0.0$~V, and is given for comparison. Dashed lines serve as guide to the eye.} 
	\label{fig:dos_tc_bias1}
\end{figure*}

Tabulated values of $\lambda_{\text{p}}$ for different charge carrier concentrations is given in Supplemental Material (SM), Tables~S1 and~S2.
It is seen, that main contribution in case of electron-doping arises from acoustic phonons, for both $n$ and $p$-doping cases.
However, for $N_{e}<$~\con{2}{14} individual contributions of acoustic and optical phonons are of the same order.
Distribution of the nesting function in ${\bf q}$-space for $N_{\text{e}}=$~\con{1}{15} is more complex [Fig.~\ref{fig:lambda_single}(b)] than for the hole-doping [Fig.~\ref{fig:lambda_single}(a)]. This is because electron scattering now involves considerable intraband transitions.
Therefore, it is not possible to determine specific scattering directions in this case. Here, the dominant contribution to $\lambda$ arises from small ${\bf q}$ (see SM, Fig.~S2), contrary to a short-wavelength character of the electron-phonon coupling at $N_{\text{h}}=$~\con{3.7}{14}.

\subsection{Effects of bias voltage}

\begin{figure*}[ht]
	\begin{minipage}[c]{8.0cm}
		\centering
		\includegraphics[width=\textwidth]{{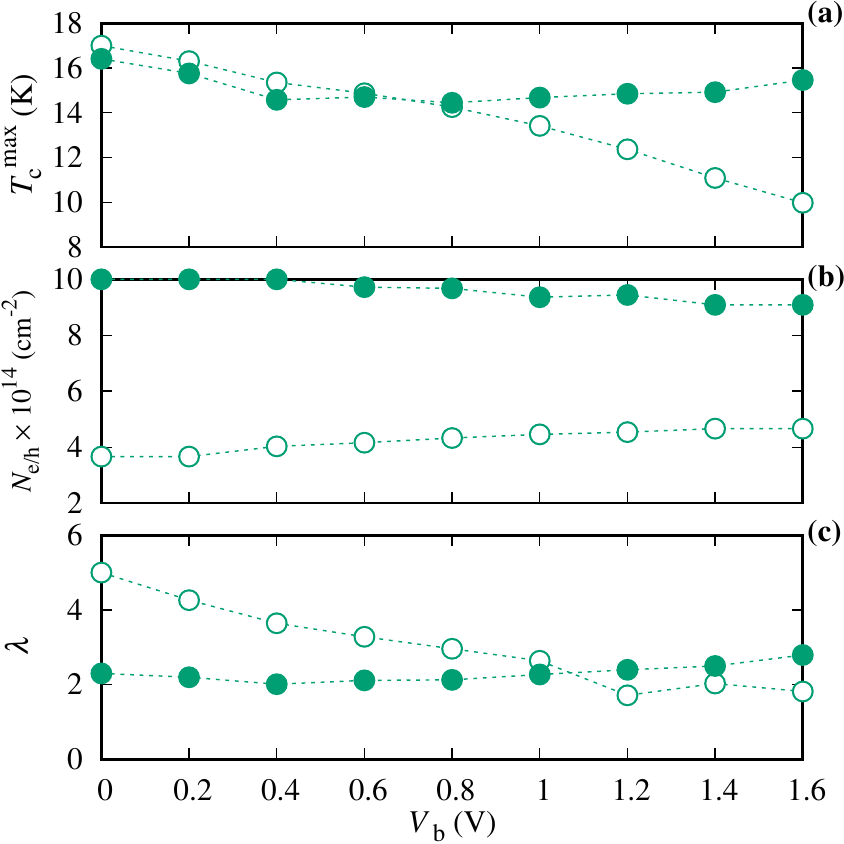}} 
	\end{minipage}\hspace{0.1cm}
	\begin{minipage}[c]{8.0cm}	
		\includegraphics[width=\textwidth]{{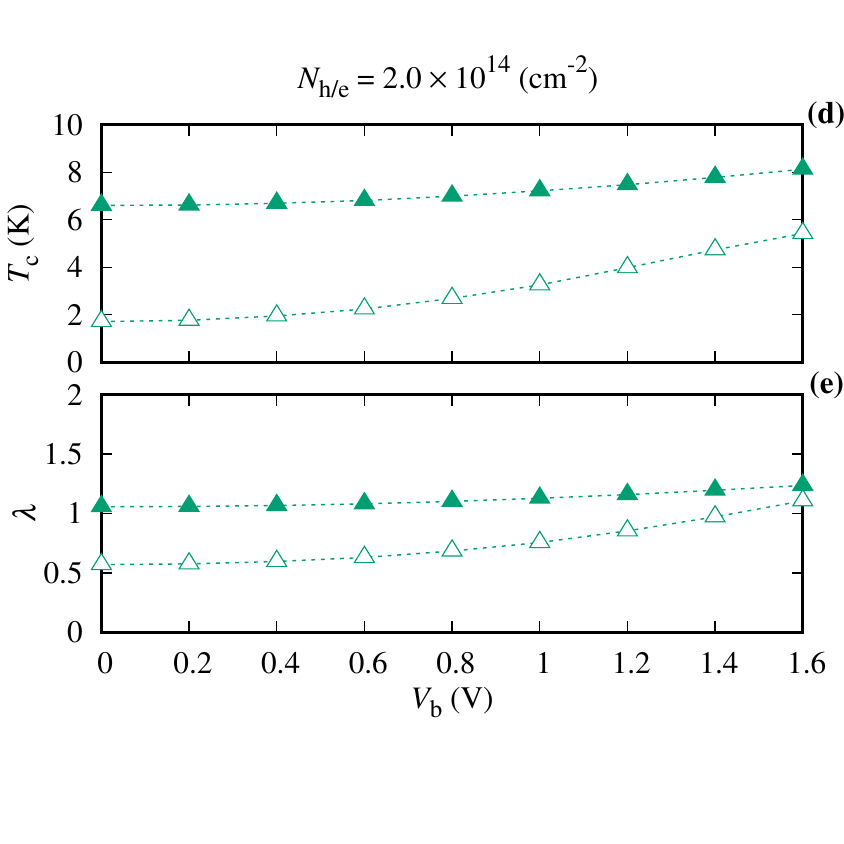}}
	\end{minipage}
	\caption{Dependence of critical temperature and electron-phonon coupling in doped antimonene on the bias voltage. Filled and open symbols correspond to $n$- and $p$-doping cases: (a) maximal critical temperature $T_{\text{c}}^{\text{max}}$; (b) charge carrier concentration $N_{\text{e/h}}$ corresponding to $T_{\text{c}}^{\text{max}}$; (c) total electron-phonon coupling strength $\lambda$. (d) and (e) show variation of critical temperature and electron-phonon coupling strength with bias voltage at fixed concentrations  $N_{\text{e/h}}=$~\con{2}{14}. }
	\label{fig:tcmax}
\end{figure*}

The electronic bands of pristine antimonene are doubly degenerate with respect to spin~\cite{Rudenko2017}, which is governed by the inversion symmetry.
When an electric field is applied, the inversion symmetry is broken and the spin degeneracy is lifted, which gives rise to the band splitting~\cite{Prishchenko2018}.
This can be clearly seen from Fig.~\ref{fig:dos_bias}, where we show the effect of bias voltage on the electronic structure of antimonene.
Apart from the SOC related splitting, the band gap is enhanced by the field, unlike, for example, few-layer phosphorene~\cite{Kim723,Liu2015,Rudenko2015}, where an opposite trend is observed.
A small gap change of 0.1 eV is observed already at \vb=1.0~V, while the highest considered bias voltage of 1.6 V results in a gap of around 30\% larger than the original one. DOS changes accordingly, as one can see from Fig.~\ref{fig:dos_bias}(a).
Taking into account strong correlation of $\lambda$ with DOS at the Fermi level, bias voltage is expected to have an effect on the superconducting critical temperatures.
DOS at $n$- and $p$-doping behaves differently with the application of electric field. The most significant change of \tc~is expected for $p$-doping at the concentrations corresponding to VHS at \vb~$=$~0~V.
Because of the strong band splitting in the $\Gamma$--\textit{K} direction, the flat band at $\varepsilon_F\approx -1.3$~eV splits, resulting in two separate peaks in DOS. In contrast to the original VHS, the two resulting peaks have significantly smaller DOS, which becomes more clear at larger voltages.
At the same time, DOS at lower charge carrier concentrations slowly increases with \vb, which is mostly observed for $p$-doping. The splitting of the VHS in the conduction band ($n$-doping) is less prominent.

Let us now consider how \tc~depends on the charge carrier concentration in the presence of \vb~$=1.0$~V.
At this voltage one can clearly see qualitative changes of the carrier DOS [Fig.~\ref{fig:dos_bias}(a)], i.e. the splitting of a peak at $\varepsilon_F\approx -1.3$~eV. 
As can be seen from Fig.~\ref{fig:dos_tc_bias1}(b), the change of \tc~for the electron-doping is nearly negligible at concentrations $N_{\text{e}}<$~\con{3}{14}. At higher concentrations, \tc~increase of the order of 1.5 K is observed, with the exception of the highest considered concentration, which results in $T_\mathrm{c}$ being 1.5~K smaller than in the \vb~$=0$~V case.
Maximum achievable \tc~is now 14.6 K for high electron concentration \con{9.4}{14}, while hole-doping gives comparable values of 13.4 and 12.1~K at $N_{\text{h}}$ of \pten{2.7}{14}~and \con{4.5}{14}, respectively. This corresponds to two new local maxima in DOS. Interestingly, \tc~at these points are in good agreement with those observed at the same concentrations at \vb$~=~$0~V.

High holes concentration of~\con{8.4}{14}~yields the critical temperature of 13~K, approximately 3~K higher than the highest considered concentration of~\con{1}{15}~in the absence of the bias voltage. 
Lower holes concentrations also yield higher \tc~than in the absence of electric field: \con{2}{14} now gives \tc~of 3.2~K, which is nearly twice as large as it was at \vb~$=0$~V.
While the absolute value of \tc~enhancement at this concentration is small, it represents a practically important trend: Increase of the bias voltage allows us to achieve higher values of \tc~at lower concentrations.

We now turn to the dependencies of maximal critical temperature $T_{\text{c}}^{\text{max}}$ on the bias voltage [Fig.~\ref{fig:tcmax}(a)] in the chosen charge carrier concentration range.
As expected, the observed trends are different for different doping types. In the case of $n$-doping, $T_{\text{c}}^{\text{max}}$ does not change significantly with bias, ranging from 14.6 to 16.4 K with the lowest value corresponding to \vb~$=0.4$~V.   
Concentrations required to achieve $T_{\text{c}}^{\text{max}}$ slowly decrease with bias voltage yet remain high: $N_{\text{e}}=$~\con{9}{14} at \vb~$=1.6$~V. 
As shown in Fig.~\ref{fig:tcmax}(c), strong electron phonon coupling is observed for all considered concentrations with $\lambda$ in the range of 2--3. At low bias voltages, $\lambda$ decreases first until \vb~$=0.4$ V and then increases in the rest of the range. The opposite trend is observed for $p$-doping.
In this case, $T_{\text{c}}^{\text{max}}$ decreases monotonously with \vb~from 17 to 10~K.
The required hole concentration, on the contrary, increases, but does not exceed \con{4.7}{14}, which is still significantly smaller than for electrons. In the presence of bias voltage, $\lambda$ reduces significantly from 5 to 1.8, indicating a sufficiently strong electron-phonon coupling.

Charge carrier concentrations presented in Fig.~\ref{fig:tcmax}(b) are high, 
and thus may be difficult to achieve in practice. At the same time, 
bias voltage could increase DOS, as well as $\lambda$ and \tc~at moderate concentrations, i.e., below \con{4}{14}.
In Figs.~\ref{fig:tcmax}(d)-\ref{fig:tcmax}(e), we consider $\lambda$ and \tc~at more realistic concentrations for both doping cases $N=$~\con{2}{14}, which are typical values in the medium concentration range. In case of electron-doping, both $\lambda$ and \tc~increase monotonously with \vb, yet for the highest bias voltage considered, \tc~increases by only $\approx$1~K.
Hole-doping yields $\approx$2 K enhancement of \tc~already at \vb~$=1.0$~V, while $\lambda$ reaches the value of 2.0.
In contrast to the case of electrons, $\lambda$ and \tc~exhibit a maximum at \vb~$\approx1.2$ V, after which one can see a gradual decrease of their values. This behavior can be attributed to a decreasing DOS, shown in Fig.~\ref{fig:dos_bias}.
Although in absolute values the effect of bias voltage is not large, it provides a possibility to
increase \tc~and $\lambda$ in antimonene by up to 20\%.

	\subsection{Dynamical stability of doped antimonene}\label{subsec:stability}

	In this section, we discuss the dynamical stability of doped antimonene.
    To model the doping effect beyond the rigid band approximation, we use the jellium doping technique.
    We are interested in studying the stability at doping concentrations corresponding to the strong Fermi-nesting, considered in the previous section, as well as to the highest charge carrier concentrations in case of $n$-doping. 
    In both cases the loss of structural stability would naturally provide the limiting factor for the applicability of used approximations.

\begin{figure}
		\includegraphics[]{{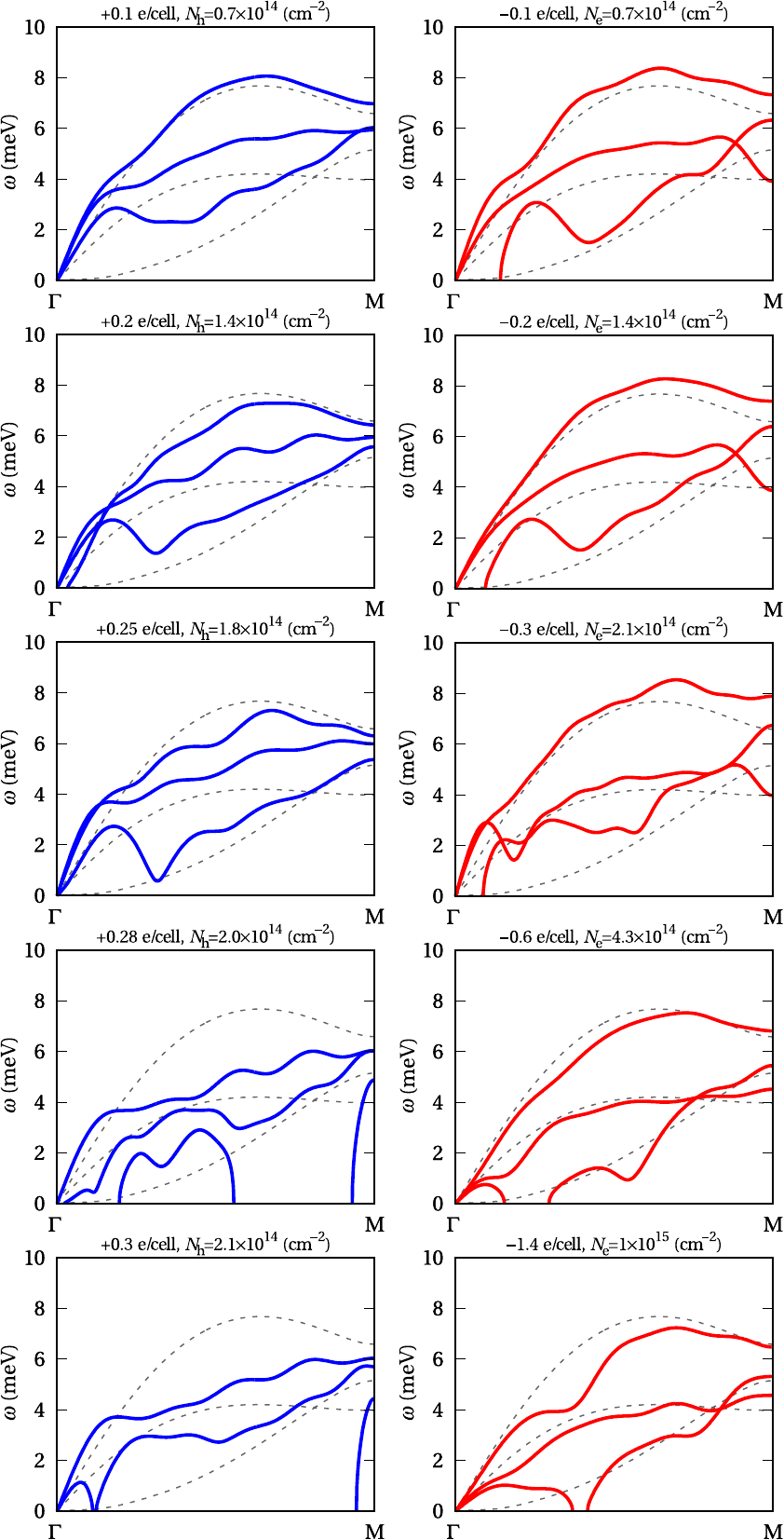}}
		\caption{Phonon spectra of hole- and electron-doped antimonene (left and right columns respectively) in $\Gamma$--M high-symmetry direction at various charge carrier concentrations, simulated using jellium doping technique. Dashed line corresponds to phonon spectra in undoped case.}\label{fig:fig9}
	\end{figure}

The results of phonon spectra calculations are presented in Fig.~\ref{fig:fig9}.
For brevity, we consider the $\Gamma$--\textit{M} direction only. Qualitatively similar effects are observed in the $\Gamma$--\textit{K} direction.
Let us first consider the hole-doping case. At low enough doping, the system demonstrates the enhancement of acoustic phonon frequencies in the long-wavelength limit, which is especially evident for TA mode. At the same time, out-of-plane phonon mode ZA becomes nearly linear at low ${\bf q}$.
	These effects persist in the concentration range from~\con{0.7}{14} to~\con{1.8}{14}.
	At~\con{1.8}{14}, a pronounced softening of ZA mode in the $\Gamma$--\textit{M} direction is observed. At the concentration of~\con{2.0}{14}, this mode becomes imaginary in a wide range of ${\bf q}$, including the vicinity of high-symmetry \textit{M} point. The TA mode also demonstrates pronounced softening.
	Further imaginary frequencies appear for the TA mode at a slightly higher hole concentration of~\con{2.1}{14} (+0.3 e/cell).
	It is important to distinguish out-of-plane mode related instability and in-plane instability.
	The appearance of the former is reported for other doped 2D materials~\cite{Margine2014,Kong2018} and is not considered as a sign of instability in the case of realistic applications, mainly due to the presence of the substrate in experimental setup. 
	At the same time, the presence of in-plane instability is an explicit indication of instability. 
	Therefore, we conclude that antimonene at concentrations $N_{\text{h}}>$~\con{2}{14} is structurally unstable.
	It is worth mentioning that to a certain extent strain can be used to stabilize the system~\cite{Zeng2016}, even if in-plane instability appears in the phonon spectrum.

On the contrary, electron-doping would only lead to out-of-plane instability, even at the highest concentration considered. 
Unlike the $p$-doping case, electron-doped antimonene is stable at~\con{2}{14} and higher. At lower concentration considered, i.e.,~\con{0.7}{14}, only the ZA mode demonstrates imaginary frequencies at low ${\bf q}$, while in-plane modes demonstrate stiffening. 
At~\con{2.1}{14}, however, softening of TA mode is observed, which further increases at higher concentrations
The doping effect on longitudinal modes is not as pronounced, with significant softening taking place only at highest charge carrier concentrations considered. This behavior is different from the hole-doping case, where the LA frequencies decrease significantly, especially away from the zone center.  
The reason for the higher stability of electron-doped antimonene can partially be associated with a more distributed electron nesting function [Fig.~\ref{fig:lambda_contrib_nza}(a)], as well as $\lambda$~(Fig.~S1 in Supplemental Materials).

	Let us now discuss possible effect of bias voltage on the stability and \tc~in light of the results presented above. 
	First of all, in the case of hole-doping, the effect of bias voltage on the electronic structure remains an important factor, since the increase of \tc~is already observed at $N_{\text{h}}$=~\con{2}{14}, where the system remains stable.
	Furthermore, the splitting of the DOS peak in the presence of $V_{\text{b}}$ allows us to assume that bias voltage may actually increase the stability of doped antimonene at $N_{\text{h}}>$~\con{2}{14} due to the modification of DOS and suppression of VHS (see Fig.~\ref{fig:dos_bias}).
    While the effect of bias voltage on the electronic structure of $n$-doped antimonene is less prominent, we expect the tendencies described for this case in the previous section to remain.

\FloatBarrier
\section{Summary and Conclusion}

We performed \textit{ab initio} calculations of electron-phonon coupling for $n$- and $p$-doped antimonene using state-of-the-art computational techniques.
We estimated critical temperature of the superconducting transition at various charge carrier concentrations in a wide range from \pten{5}{13} to \con{1}{15}~using the Allen-Dynes-McMillan equation.
{\color{black}Doing so, we considered the dynamical stability of studied configurations.}

Dependence of the electron-phonon coupling strength from charge carrier concentration is essentially conditioned by the doping type and is mainly determined by the Fermi surface topologies.
We find that at hole concentrations below \con{2}{14} and electron concentration below \con{5}{13} superconductivity is not expected above 0.5~K. 
Higher charge carrier concentrations yield \tc~in a wide range.
{\color{black}However, the increase of charge carrier concentration in the case of hole-doping is limited by the dynamical instability, which occurs for $N_{\text{h}}>$\con{2.0}{14}. This behavior can be explained by the presence of van Hove singularity close to dynamical stability limit in the density of hole states. Among the stable configurations, the maximum value of~\tc~is estimated to be 15~K, and corresponds to the electron-doping case. The hole-doping yields $T_{\text{c}}^{\text{max}}\approx1.5$~K at the concentrations in the range of 1.8--\con{2.0}{14}.}
These value are comparable or exceed those reported for other doped elemental monolayer materials.
The value of strong electron-phonon coupling strength as high as $\lambda=2.3$ is predicted for the highest considered electron concentration of \con{1}{15}.
For all studied charge carrier concentrations the main contribution to electron-phonon coupling arises from the interaction of carriers with in-plane acoustic phonons.

We also studied the effects of electric field applied in the direction perpendicular to the atomic sheet, on electron-phonon coupling and critical temperature of antimonene.
In the bias voltage range of 0.0--1.6~V, the electronic structure of antimonene undergoes spin-orbit-assisted bands splitting, as well as enhancement of the band gap width. 
This effect allows us to increase the \tc~for the stable hole-doped configurations; thus, in the vicinity of destabilization point at~\con{2}{14} the maximum increase of \tc~reaches 3.5 K at the highest considered bias voltage.
In contrast, the electron-doping case is less affected by the application of bias voltage, leading to a slight variation of the critical temperature and electron-phonon coupling over the whole concentration range, which potentially simplifies the application of gating-based doping methods for the system and making results more predictable.
Overall, bias voltage allows us to control electron-phonon coupling as well as related properties in heavy-element 2D semiconductors, making them interesting objects for further studies.

\section{Acknowledgments}

This work is part of the research programme ``Two-dimensional semiconductor crystals'' with project number 14TWOD01, which is (partly) financed by the Netherlands Organisation for Scientific Research (NWO).
The calculations were preformed at computational cluster ``TCM'', Radboud University, Nijmegen, Nederlands and at computational cluster NUST ``MISIS'', Moscow, Russia. A.N.R. acknowledges  support from the Russian Science Foundation, Grant No. 17-72-20041.

\nocite{apsrev41Control}
\bibliographystyle{apsrev4-1}
\bibliography{sbbib}

\begin{thebibliography}{77}%
\makeatletter
\providecommand \@ifxundefined [1]{%
 \@ifx{#1\undefined}
}%
\providecommand \@ifnum [1]{%
 \ifnum #1\expandafter \@firstoftwo
 \else \expandafter \@secondoftwo
 \fi
}%
\providecommand \@ifx [1]{%
 \ifx #1\expandafter \@firstoftwo
 \else \expandafter \@secondoftwo
 \fi
}%
\providecommand \natexlab [1]{#1}%
\providecommand \enquote  [1]{``#1''}%
\providecommand \bibnamefont  [1]{#1}%
\providecommand \bibfnamefont [1]{#1}%
\providecommand \citenamefont [1]{#1}%
\providecommand \href@noop [0]{\@secondoftwo}%
\providecommand \href [0]{\begingroup \@sanitize@url \@href}%
\providecommand \@href[1]{\@@startlink{#1}\@@href}%
\providecommand \@@href[1]{\endgroup#1\@@endlink}%
\providecommand \@sanitize@url [0]{\catcode `\\12\catcode `\$12\catcode
  `\&12\catcode `\#12\catcode `\^12\catcode `\_12\catcode `\%12\relax}%
\providecommand \@@startlink[1]{}%
\providecommand \@@endlink[0]{}%
\providecommand \url  [0]{\begingroup\@sanitize@url \@url }%
\providecommand \@url [1]{\endgroup\@href {#1}{\urlprefix }}%
\providecommand \urlprefix  [0]{URL }%
\providecommand \Eprint [0]{\href }%
\providecommand \doibase [0]{http://dx.doi.org/}%
\providecommand \selectlanguage [0]{\@gobble}%
\providecommand \bibinfo  [0]{\@secondoftwo}%
\providecommand \bibfield  [0]{\@secondoftwo}%
\providecommand \translation [1]{[#1]}%
\providecommand \BibitemOpen [0]{}%
\providecommand \bibitemStop [0]{}%
\providecommand \bibitemNoStop [0]{.\EOS\space}%
\providecommand \EOS [0]{\spacefactor3000\relax}%
\providecommand \BibitemShut  [1]{\csname bibitem#1\endcsname}%
\let\auto@bib@innerbib\@empty
\bibitem [{\citenamefont {Zhang}\ \emph {et~al.}(2015)\citenamefont {Zhang},
  \citenamefont {Yan}, \citenamefont {Li}, \citenamefont {Chen},\ and\
  \citenamefont {Zeng}}]{Zhang2015}%
  \BibitemOpen
  \bibfield  {author} {\bibinfo {author} {\bibfnamefont {S.}~\bibnamefont
  {Zhang}}, \bibinfo {author} {\bibfnamefont {Z.}~\bibnamefont {Yan}}, \bibinfo
  {author} {\bibfnamefont {Y.}~\bibnamefont {Li}}, \bibinfo {author}
  {\bibfnamefont {Z.}~\bibnamefont {Chen}}, \ and\ \bibinfo {author}
  {\bibfnamefont {H.}~\bibnamefont {Zeng}},\ }\bibfield  {title} {\enquote
  {\bibinfo {title} {{Atomically Thin Arsenene and Antimonene:
  Semimetal-Semiconductor and Indirect-Direct Band-Gap Transitions}},}\ }\href
  {\doibase 10.1002/anie.201411246} {\bibfield  {journal} {\bibinfo  {journal}
  {Angew. Chem. Int. Ed.}\ }\textbf {\bibinfo {volume} {54}},\ \bibinfo {pages}
  {3112} (\bibinfo {year} {2015})}\BibitemShut {NoStop}%
\bibitem [{\citenamefont {Lei}\ \emph {et~al.}(2016)\citenamefont {Lei},
  \citenamefont {Liu}, \citenamefont {Zhao}, \citenamefont {Li}, \citenamefont
  {Li}, \citenamefont {Wang}, \citenamefont {Wu}, \citenamefont {Qian},
  \citenamefont {Wang},\ and\ \citenamefont {Ibrahim}}]{Lei2016}%
  \BibitemOpen
  \bibfield  {author} {\bibinfo {author} {\bibfnamefont {T.}~\bibnamefont
  {Lei}}, \bibinfo {author} {\bibfnamefont {C.}~\bibnamefont {Liu}}, \bibinfo
  {author} {\bibfnamefont {J.-L.}\ \bibnamefont {Zhao}}, \bibinfo {author}
  {\bibfnamefont {J.-M.}\ \bibnamefont {Li}}, \bibinfo {author} {\bibfnamefont
  {Y.-P.}\ \bibnamefont {Li}}, \bibinfo {author} {\bibfnamefont {J.-O.}\
  \bibnamefont {Wang}}, \bibinfo {author} {\bibfnamefont {R.}~\bibnamefont
  {Wu}}, \bibinfo {author} {\bibfnamefont {H.-J.}\ \bibnamefont {Qian}},
  \bibinfo {author} {\bibfnamefont {H.-Q.}\ \bibnamefont {Wang}}, \ and\
  \bibinfo {author} {\bibfnamefont {K.}~\bibnamefont {Ibrahim}},\ }\bibfield
  {title} {\enquote {\bibinfo {title} {{Electronic structure of antimonene
  grown on Sb$_{2}$Te$_3$ (111) and Bi$_2$Te$_3$ substrates}},}\ }\href
  {\doibase 10.1063/1.4939281} {\bibfield  {journal} {\bibinfo  {journal} {J.
  Appl. Phys.}\ }\textbf {\bibinfo {volume} {119}},\ \bibinfo {pages} {015302}
  (\bibinfo {year} {2016})}\BibitemShut {NoStop}%
\bibitem [{\citenamefont {Vogt}\ \emph {et~al.}(2012)\citenamefont {Vogt},
  \citenamefont {{De Padova}}, \citenamefont {Quaresima}, \citenamefont
  {Avila}, \citenamefont {Frantzeskakis}, \citenamefont {Asensio},
  \citenamefont {Resta}, \citenamefont {Ealet},\ and\ \citenamefont {{Le
  Lay}}}]{Vogt2012}%
  \BibitemOpen
  \bibfield  {author} {\bibinfo {author} {\bibfnamefont {P.}~\bibnamefont
  {Vogt}}, \bibinfo {author} {\bibfnamefont {P.}~\bibnamefont {{De Padova}}},
  \bibinfo {author} {\bibfnamefont {C.}~\bibnamefont {Quaresima}}, \bibinfo
  {author} {\bibfnamefont {J.}~\bibnamefont {Avila}}, \bibinfo {author}
  {\bibfnamefont {E.}~\bibnamefont {Frantzeskakis}}, \bibinfo {author}
  {\bibfnamefont {M.~C.}\ \bibnamefont {Asensio}}, \bibinfo {author}
  {\bibfnamefont {A.}~\bibnamefont {Resta}}, \bibinfo {author} {\bibfnamefont
  {B.}~\bibnamefont {Ealet}}, \ and\ \bibinfo {author} {\bibfnamefont
  {G.}~\bibnamefont {{Le Lay}}},\ }\bibfield  {title} {\enquote {\bibinfo
  {title} {{Silicene: Compelling Experimental Evidence for Graphenelike
  Two-Dimensional Silicon}},}\ }\href {\doibase 10.1103/PhysRevLett.108.155501}
  {\bibfield  {journal} {\bibinfo  {journal} {Phys. Rev. Lett.}\ }\textbf
  {\bibinfo {volume} {108}},\ \bibinfo {pages} {155501} (\bibinfo {year}
  {2012})}\BibitemShut {NoStop}%
\bibitem [{\citenamefont {Zhang}\ \emph {et~al.}(2016)\citenamefont {Zhang},
  \citenamefont {Bampoulis}, \citenamefont {Rudenko}, \citenamefont {Yao},
  \citenamefont {van Houselt}, \citenamefont {Poelsema}, \citenamefont
  {Katsnelson},\ and\ \citenamefont {Zandvliet}}]{Zhang2016}%
  \BibitemOpen
  \bibfield  {author} {\bibinfo {author} {\bibfnamefont {L.}~\bibnamefont
  {Zhang}}, \bibinfo {author} {\bibfnamefont {P.}~\bibnamefont {Bampoulis}},
  \bibinfo {author} {\bibfnamefont {A.~N.}\ \bibnamefont {Rudenko}}, \bibinfo
  {author} {\bibfnamefont {Q.}~\bibnamefont {Yao}}, \bibinfo {author}
  {\bibfnamefont {A.}~\bibnamefont {van Houselt}}, \bibinfo {author}
  {\bibfnamefont {B.}~\bibnamefont {Poelsema}}, \bibinfo {author}
  {\bibfnamefont {M.~I.}\ \bibnamefont {Katsnelson}}, \ and\ \bibinfo {author}
  {\bibfnamefont {H.~J.~W.}\ \bibnamefont {Zandvliet}},\ }\bibfield  {title}
  {\enquote {\bibinfo {title} {{Structural and Electronic Properties of
  Germanene on MoS$_{2}$}},}\ }\href {\doibase 10.1103/PhysRevLett.116.256804}
  {\bibfield  {journal} {\bibinfo  {journal} {Phys. Rev. Lett.}\ }\textbf
  {\bibinfo {volume} {116}},\ \bibinfo {pages} {256804} (\bibinfo {year}
  {2016})}\BibitemShut {NoStop}%
\bibitem [{\citenamefont {Gu}\ \emph {et~al.}(2017)\citenamefont {Gu},
  \citenamefont {Zhao}, \citenamefont {Zhang}, \citenamefont {Sun},
  \citenamefont {Yuan}, \citenamefont {Hu}, \citenamefont {Han}, \citenamefont
  {Ma}, \citenamefont {Wang}, \citenamefont {Huo}, \citenamefont {Huang},
  \citenamefont {Li},\ and\ \citenamefont {Chen}}]{Gu2017}%
  \BibitemOpen
  \bibfield  {author} {\bibinfo {author} {\bibfnamefont {C.}~\bibnamefont
  {Gu}}, \bibinfo {author} {\bibfnamefont {S.}~\bibnamefont {Zhao}}, \bibinfo
  {author} {\bibfnamefont {J.~L.}\ \bibnamefont {Zhang}}, \bibinfo {author}
  {\bibfnamefont {S.}~\bibnamefont {Sun}}, \bibinfo {author} {\bibfnamefont
  {K.}~\bibnamefont {Yuan}}, \bibinfo {author} {\bibfnamefont {Z.}~\bibnamefont
  {Hu}}, \bibinfo {author} {\bibfnamefont {C.}~\bibnamefont {Han}}, \bibinfo
  {author} {\bibfnamefont {Z.}~\bibnamefont {Ma}}, \bibinfo {author}
  {\bibfnamefont {L.}~\bibnamefont {Wang}}, \bibinfo {author} {\bibfnamefont
  {F.}~\bibnamefont {Huo}}, \bibinfo {author} {\bibfnamefont {W.}~\bibnamefont
  {Huang}}, \bibinfo {author} {\bibfnamefont {Z.}~\bibnamefont {Li}}, \ and\
  \bibinfo {author} {\bibfnamefont {W.}~\bibnamefont {Chen}},\ }\bibfield
  {title} {\enquote {\bibinfo {title} {{Growth of Quasi-Free-Standing
  Single-Layer Blue Phosphorus on Tellurium Monolayer Functionalized
  Au(111)}},}\ }\href {\doibase 10.1021/acsnano.7b01575} {\bibfield  {journal}
  {\bibinfo  {journal} {ACS Nano}\ }\textbf {\bibinfo {volume} {11}},\ \bibinfo
  {pages} {4943} (\bibinfo {year} {2017})}\BibitemShut {NoStop}%
\bibitem [{\citenamefont {Ares}\ \emph {et~al.}(2018)\citenamefont {Ares},
  \citenamefont {Palacios}, \citenamefont {Abell{\'{a}}n}, \citenamefont
  {G{\'{o}}mez-Herrero},\ and\ \citenamefont {Zamora}}]{Ares2018}%
  \BibitemOpen
  \bibfield  {author} {\bibinfo {author} {\bibfnamefont {P.}~\bibnamefont
  {Ares}}, \bibinfo {author} {\bibfnamefont {J.~J.}\ \bibnamefont {Palacios}},
  \bibinfo {author} {\bibfnamefont {G.}~\bibnamefont {Abell{\'{a}}n}}, \bibinfo
  {author} {\bibfnamefont {J.}~\bibnamefont {G{\'{o}}mez-Herrero}}, \ and\
  \bibinfo {author} {\bibfnamefont {F.}~\bibnamefont {Zamora}},\ }\bibfield
  {title} {\enquote {\bibinfo {title} {{Recent Progress on Antimonene: A New
  Bidimensional Material}},}\ }\href {\doibase 10.1002/adma.201703771}
  {\bibfield  {journal} {\bibinfo  {journal} {Adv. Mater.}\ }\textbf {\bibinfo
  {volume} {30}},\ \bibinfo {pages} {1703771} (\bibinfo {year}
  {2018})}\BibitemShut {NoStop}%
\bibitem [{\citenamefont {Wu}\ \emph {et~al.}(2017)\citenamefont {Wu},
  \citenamefont {Shao}, \citenamefont {Liu}, \citenamefont {Feng},
  \citenamefont {Wang}, \citenamefont {Sun}, \citenamefont {Liu}, \citenamefont
  {Wang}, \citenamefont {Liu}, \citenamefont {Zhu}, \citenamefont {Wang},
  \citenamefont {Du}, \citenamefont {Shi}, \citenamefont {Ibrahim},\ and\
  \citenamefont {Gao}}]{Wu2017}%
  \BibitemOpen
  \bibfield  {author} {\bibinfo {author} {\bibfnamefont {X.}~\bibnamefont
  {Wu}}, \bibinfo {author} {\bibfnamefont {Y.}~\bibnamefont {Shao}}, \bibinfo
  {author} {\bibfnamefont {H.}~\bibnamefont {Liu}}, \bibinfo {author}
  {\bibfnamefont {Z.}~\bibnamefont {Feng}}, \bibinfo {author} {\bibfnamefont
  {Y.~L.}\ \bibnamefont {Wang}}, \bibinfo {author} {\bibfnamefont {J.~T.}\
  \bibnamefont {Sun}}, \bibinfo {author} {\bibfnamefont {C.}~\bibnamefont
  {Liu}}, \bibinfo {author} {\bibfnamefont {J.~O.}\ \bibnamefont {Wang}},
  \bibinfo {author} {\bibfnamefont {Z.~L.}\ \bibnamefont {Liu}}, \bibinfo
  {author} {\bibfnamefont {S.~Y.}\ \bibnamefont {Zhu}}, \bibinfo {author}
  {\bibfnamefont {Y.~Q.}\ \bibnamefont {Wang}}, \bibinfo {author}
  {\bibfnamefont {S.~X.}\ \bibnamefont {Du}}, \bibinfo {author} {\bibfnamefont
  {Y.~G.}\ \bibnamefont {Shi}}, \bibinfo {author} {\bibfnamefont
  {K.}~\bibnamefont {Ibrahim}}, \ and\ \bibinfo {author} {\bibfnamefont
  {H.~J.}\ \bibnamefont {Gao}},\ }\bibfield  {title} {\enquote {\bibinfo
  {title} {{Epitaxial Growth and Air-Stability of Monolayer Antimonene on
  PdTe$_2$}},}\ }\href {\doibase 10.1002/adma.201605407} {\bibfield  {journal}
  {\bibinfo  {journal} {Adv. Mater.}\ }\textbf {\bibinfo {volume} {29}},\
  \bibinfo {pages} {1605407} (\bibinfo {year} {2017})}\BibitemShut {NoStop}%
\bibitem [{\citenamefont {Gibaja}\ \emph {et~al.}(2016)\citenamefont {Gibaja},
  \citenamefont {Rodriguez-San-Miguel}, \citenamefont {Ares}, \citenamefont
  {G{\'{o}}mez-Herrero}, \citenamefont {Varela}, \citenamefont {Gillen},
  \citenamefont {Maultzsch}, \citenamefont {Hauke}, \citenamefont {Hirsch},
  \citenamefont {Abell{\'{a}}n},\ and\ \citenamefont {Zamora}}]{Gibaja2016}%
  \BibitemOpen
  \bibfield  {author} {\bibinfo {author} {\bibfnamefont {C.}~\bibnamefont
  {Gibaja}}, \bibinfo {author} {\bibfnamefont {D.}~\bibnamefont
  {Rodriguez-San-Miguel}}, \bibinfo {author} {\bibfnamefont {P.}~\bibnamefont
  {Ares}}, \bibinfo {author} {\bibfnamefont {J.}~\bibnamefont
  {G{\'{o}}mez-Herrero}}, \bibinfo {author} {\bibfnamefont {M.}~\bibnamefont
  {Varela}}, \bibinfo {author} {\bibfnamefont {R.}~\bibnamefont {Gillen}},
  \bibinfo {author} {\bibfnamefont {J.}~\bibnamefont {Maultzsch}}, \bibinfo
  {author} {\bibfnamefont {F.}~\bibnamefont {Hauke}}, \bibinfo {author}
  {\bibfnamefont {A.}~\bibnamefont {Hirsch}}, \bibinfo {author} {\bibfnamefont
  {G.}~\bibnamefont {Abell{\'{a}}n}}, \ and\ \bibinfo {author} {\bibfnamefont
  {F.}~\bibnamefont {Zamora}},\ }\bibfield  {title} {\enquote {\bibinfo {title}
  {{Few-Layer Antimonene by Liquid-Phase Exfoliation}},}\ }\href {\doibase
  10.1002/anie.201605298} {\bibfield  {journal} {\bibinfo  {journal} {Angew.
  Chem. Int. Ed.}\ }\textbf {\bibinfo {volume} {55}},\ \bibinfo {pages} {14345}
  (\bibinfo {year} {2016})}\BibitemShut {NoStop}%
\bibitem [{\citenamefont {Ares}\ \emph {et~al.}(2016)\citenamefont {Ares},
  \citenamefont {Aguilar-Galindo}, \citenamefont {Rodr{\'{i}}guez-San-Miguel},
  \citenamefont {Aldave}, \citenamefont {D{\'{i}}az-Tendero}, \citenamefont
  {Alcam{\'{i}}}, \citenamefont {Mart{\'{i}}n}, \citenamefont
  {G{\'{o}}mez-Herrero},\ and\ \citenamefont {Zamora}}]{Ares2016}%
  \BibitemOpen
  \bibfield  {author} {\bibinfo {author} {\bibfnamefont {P.}~\bibnamefont
  {Ares}}, \bibinfo {author} {\bibfnamefont {F.}~\bibnamefont
  {Aguilar-Galindo}}, \bibinfo {author} {\bibfnamefont {D.}~\bibnamefont
  {Rodr{\'{i}}guez-San-Miguel}}, \bibinfo {author} {\bibfnamefont {D.~A.}\
  \bibnamefont {Aldave}}, \bibinfo {author} {\bibfnamefont {S.}~\bibnamefont
  {D{\'{i}}az-Tendero}}, \bibinfo {author} {\bibfnamefont {M.}~\bibnamefont
  {Alcam{\'{i}}}}, \bibinfo {author} {\bibfnamefont {F.}~\bibnamefont
  {Mart{\'{i}}n}}, \bibinfo {author} {\bibfnamefont {J.}~\bibnamefont
  {G{\'{o}}mez-Herrero}}, \ and\ \bibinfo {author} {\bibfnamefont
  {F.}~\bibnamefont {Zamora}},\ }\bibfield  {title} {\enquote {\bibinfo {title}
  {{Mechanical Isolation of Highly Stable Antimonene under Ambient
  Conditions}},}\ }\href {\doibase 10.1002/adma.201602128} {\bibfield
  {journal} {\bibinfo  {journal} {Adv. Mater.}\ }\textbf {\bibinfo {volume}
  {28}},\ \bibinfo {pages} {6332} (\bibinfo {year} {2016})}\BibitemShut
  {NoStop}%
\bibitem [{\citenamefont {Morishita}\ and\ \citenamefont
  {Spencer}(2015)}]{Morishita2015}%
  \BibitemOpen
  \bibfield  {author} {\bibinfo {author} {\bibfnamefont {T.}~\bibnamefont
  {Morishita}}\ and\ \bibinfo {author} {\bibfnamefont {M.~J.}\ \bibnamefont
  {Spencer}},\ }\bibfield  {title} {\enquote {\bibinfo {title} {{How silicene
  on Ag(111) oxidizes: microscopic mechanism of the reaction of O$_{2}$ with
  silicene}},}\ }\href {\doibase 10.1038/srep17570} {\bibfield  {journal}
  {\bibinfo  {journal} {Sci. Rep.}\ }\textbf {\bibinfo {volume} {5}},\ \bibinfo
  {pages} {17570} (\bibinfo {year} {2015})}\BibitemShut {NoStop}%
\bibitem [{\citenamefont {Kuriakose}\ \emph {et~al.}(2018)\citenamefont
  {Kuriakose}, \citenamefont {Ahmed}, \citenamefont {Balendhran}, \citenamefont
  {Bansal}, \citenamefont {Sriram}, \citenamefont {Bhaskaran},\ and\
  \citenamefont {Walia}}]{Kuriakose2018}%
  \BibitemOpen
  \bibfield  {author} {\bibinfo {author} {\bibfnamefont {S.}~\bibnamefont
  {Kuriakose}}, \bibinfo {author} {\bibfnamefont {T.}~\bibnamefont {Ahmed}},
  \bibinfo {author} {\bibfnamefont {S.}~\bibnamefont {Balendhran}}, \bibinfo
  {author} {\bibfnamefont {V.}~\bibnamefont {Bansal}}, \bibinfo {author}
  {\bibfnamefont {S.}~\bibnamefont {Sriram}}, \bibinfo {author} {\bibfnamefont
  {M.}~\bibnamefont {Bhaskaran}}, \ and\ \bibinfo {author} {\bibfnamefont
  {S.}~\bibnamefont {Walia}},\ }\bibfield  {title} {\enquote {\bibinfo {title}
  {{Black phosphorus: ambient degradation and strategies for protection}},}\
  }\href {\doibase 10.1088/2053-1583/aab810} {\bibfield  {journal} {\bibinfo
  {journal} {2D Mater.}\ }\textbf {\bibinfo {volume} {5}},\ \bibinfo {pages}
  {032001} (\bibinfo {year} {2018})}\BibitemShut {NoStop}%
\bibitem [{\citenamefont {Rudenko}\ \emph {et~al.}(2017)\citenamefont
  {Rudenko}, \citenamefont {Katsnelson},\ and\ \citenamefont
  {Rold{\'{a}}n}}]{Rudenko2017}%
  \BibitemOpen
  \bibfield  {author} {\bibinfo {author} {\bibfnamefont {A.~N.}\ \bibnamefont
  {Rudenko}}, \bibinfo {author} {\bibfnamefont {M.~I.}\ \bibnamefont
  {Katsnelson}}, \ and\ \bibinfo {author} {\bibfnamefont {R.}~\bibnamefont
  {Rold{\'{a}}n}},\ }\bibfield  {title} {\enquote {\bibinfo {title}
  {{Electronic properties of single-layer antimony: Tight-binding model,
  spin-orbit coupling, and the strength of effective Coulomb interactions}},}\
  }\href {\doibase 10.1103/PhysRevB.95.081407} {\bibfield  {journal} {\bibinfo
  {journal} {Phys. Rev. B}\ }\textbf {\bibinfo {volume} {95}},\ \bibinfo
  {pages} {081407} (\bibinfo {year} {2017})}\BibitemShut {NoStop}%
\bibitem [{\citenamefont {Pizzi}\ \emph {et~al.}(2016)\citenamefont {Pizzi},
  \citenamefont {Gibertini}, \citenamefont {Dib}, \citenamefont {Marzari},
  \citenamefont {Iannaccone},\ and\ \citenamefont {Fiori}}]{Pizzi2016}%
  \BibitemOpen
  \bibfield  {author} {\bibinfo {author} {\bibfnamefont {G.}~\bibnamefont
  {Pizzi}}, \bibinfo {author} {\bibfnamefont {M.}~\bibnamefont {Gibertini}},
  \bibinfo {author} {\bibfnamefont {E.}~\bibnamefont {Dib}}, \bibinfo {author}
  {\bibfnamefont {N.}~\bibnamefont {Marzari}}, \bibinfo {author} {\bibfnamefont
  {G.}~\bibnamefont {Iannaccone}}, \ and\ \bibinfo {author} {\bibfnamefont
  {G.}~\bibnamefont {Fiori}},\ }\bibfield  {title} {\enquote {\bibinfo {title}
  {{Performance of arsenene and antimonene double-gate MOSFETs from first
  principles}},}\ }\href {http://dx.doi.org/10.1038/ncomms12585
  10.1038/ncomms12585} {\bibfield  {journal} {\bibinfo  {journal} {Nat.
  Commun.}\ }\textbf {\bibinfo {volume} {7}},\ \bibinfo {pages} {12585}
  (\bibinfo {year} {2016})}\BibitemShut {NoStop}%
\bibitem [{\citenamefont {Singh}\ \emph {et~al.}(2016)\citenamefont {Singh},
  \citenamefont {Gupta}, \citenamefont {Sonvane},\ and\ \citenamefont
  {Luka{\v{c}}evi{\'{c}}}}]{Singh2016}%
  \BibitemOpen
  \bibfield  {author} {\bibinfo {author} {\bibfnamefont {D.}~\bibnamefont
  {Singh}}, \bibinfo {author} {\bibfnamefont {S.~K.}\ \bibnamefont {Gupta}},
  \bibinfo {author} {\bibfnamefont {Y.}~\bibnamefont {Sonvane}}, \ and\
  \bibinfo {author} {\bibfnamefont {I.}~\bibnamefont {Luka{\v{c}}evi{\'{c}}}},\
  }\bibfield  {title} {\enquote {\bibinfo {title} {{Antimonene: a monolayer
  material for ultraviolet optical nanodevices}},}\ }\href {\doibase
  10.1039/C6TC01913G} {\bibfield  {journal} {\bibinfo  {journal} {J. Mater.
  Chem. C}\ }\textbf {\bibinfo {volume} {4}},\ \bibinfo {pages} {6386}
  (\bibinfo {year} {2016})}\BibitemShut {NoStop}%
\bibitem [{\citenamefont {Zhao}\ \emph {et~al.}(2015)\citenamefont {Zhao},
  \citenamefont {Zhang},\ and\ \citenamefont {Li}}]{Zhao2015}%
  \BibitemOpen
  \bibfield  {author} {\bibinfo {author} {\bibfnamefont {M.}~\bibnamefont
  {Zhao}}, \bibinfo {author} {\bibfnamefont {X.}~\bibnamefont {Zhang}}, \ and\
  \bibinfo {author} {\bibfnamefont {L.}~\bibnamefont {Li}},\ }\bibfield
  {title} {\enquote {\bibinfo {title} {{Strain-driven band inversion and
  topological aspects in Antimonene}},}\ }\href {\doibase 10.1038/srep16108}
  {\bibfield  {journal} {\bibinfo  {journal} {Sci. Rep.}\ }\textbf {\bibinfo
  {volume} {5}},\ \bibinfo {pages} {16108} (\bibinfo {year}
  {2015})}\BibitemShut {NoStop}%
\bibitem [{\citenamefont {Uchihashi}(2017)}]{Uchihashi2017}%
  \BibitemOpen
  \bibfield  {author} {\bibinfo {author} {\bibfnamefont {T.}~\bibnamefont
  {Uchihashi}},\ }\bibfield  {title} {\enquote {\bibinfo {title}
  {{Two-dimensional superconductors with atomic-scale thickness}},}\ }\href
  {\doibase 10.1088/0953-2048/30/1/013002} {\bibfield  {journal} {\bibinfo
  {journal} {Supercond. Sci. Technol.}\ }\textbf {\bibinfo {volume} {30}},\
  \bibinfo {pages} {013002} (\bibinfo {year} {2017})}\BibitemShut {NoStop}%
\bibitem [{\citenamefont {{\"{O}}zer}\ \emph {et~al.}(2006)\citenamefont
  {{\"{O}}zer}, \citenamefont {Thompson},\ and\ \citenamefont
  {Weitering}}]{Ozer2006}%
  \BibitemOpen
  \bibfield  {author} {\bibinfo {author} {\bibfnamefont {M.~M.}\ \bibnamefont
  {{\"{O}}zer}}, \bibinfo {author} {\bibfnamefont {J.~R.}\ \bibnamefont
  {Thompson}}, \ and\ \bibinfo {author} {\bibfnamefont {H.~H.}\ \bibnamefont
  {Weitering}},\ }\bibfield  {title} {\enquote {\bibinfo {title} {{Hard
  superconductivity of a soft metal in the quantum regime}},}\ }\href {\doibase
  10.1038/nphys244} {\bibfield  {journal} {\bibinfo  {journal} {Nat. Phys.}\
  }\textbf {\bibinfo {volume} {2}},\ \bibinfo {pages} {173} (\bibinfo {year}
  {2006})}\BibitemShut {NoStop}%
\bibitem [{\citenamefont {Ludbrook}\ \emph {et~al.}(2015)\citenamefont
  {Ludbrook}, \citenamefont {Levy}, \citenamefont {Nigge}, \citenamefont
  {Zonno}, \citenamefont {Schneider}, \citenamefont {Dvorak}, \citenamefont
  {Veenstra}, \citenamefont {Zhdanovich}, \citenamefont {Wong}, \citenamefont
  {Dosanjh}, \citenamefont {Stra{\ss}er}, \citenamefont {St{\"{o}}hr},
  \citenamefont {Forti}, \citenamefont {Ast}, \citenamefont {Starke},\ and\
  \citenamefont {Damascelli}}]{Ludbrook2015}%
  \BibitemOpen
  \bibfield  {author} {\bibinfo {author} {\bibfnamefont {B.~M.}\ \bibnamefont
  {Ludbrook}}, \bibinfo {author} {\bibfnamefont {G.}~\bibnamefont {Levy}},
  \bibinfo {author} {\bibfnamefont {P.}~\bibnamefont {Nigge}}, \bibinfo
  {author} {\bibfnamefont {M.}~\bibnamefont {Zonno}}, \bibinfo {author}
  {\bibfnamefont {M.}~\bibnamefont {Schneider}}, \bibinfo {author}
  {\bibfnamefont {D.~J.}\ \bibnamefont {Dvorak}}, \bibinfo {author}
  {\bibfnamefont {C.~N.}\ \bibnamefont {Veenstra}}, \bibinfo {author}
  {\bibfnamefont {S.}~\bibnamefont {Zhdanovich}}, \bibinfo {author}
  {\bibfnamefont {D.}~\bibnamefont {Wong}}, \bibinfo {author} {\bibfnamefont
  {P.}~\bibnamefont {Dosanjh}}, \bibinfo {author} {\bibfnamefont
  {C.}~\bibnamefont {Stra{\ss}er}}, \bibinfo {author} {\bibfnamefont
  {A.}~\bibnamefont {St{\"{o}}hr}}, \bibinfo {author} {\bibfnamefont
  {S.}~\bibnamefont {Forti}}, \bibinfo {author} {\bibfnamefont {C.~R.}\
  \bibnamefont {Ast}}, \bibinfo {author} {\bibfnamefont {U.}~\bibnamefont
  {Starke}}, \ and\ \bibinfo {author} {\bibfnamefont {A.}~\bibnamefont
  {Damascelli}},\ }\bibfield  {title} {\enquote {\bibinfo {title} {{Evidence
  for superconductivity in Li-decorated monolayer graphene.}}}\ }\href
  {\doibase 10.1073/pnas.1510435112} {\bibfield  {journal} {\bibinfo  {journal}
  {Proc. Natl. Acad. Sci. U. S. A.}\ }\textbf {\bibinfo {volume} {112}},\
  \bibinfo {pages} {11795} (\bibinfo {year} {2015})}\BibitemShut {NoStop}%
\bibitem [{\citenamefont {Ge}\ \emph {et~al.}(2015)\citenamefont {Ge},
  \citenamefont {Wan}, \citenamefont {Yang},\ and\ \citenamefont
  {Yao}}]{Ge2015}%
  \BibitemOpen
  \bibfield  {author} {\bibinfo {author} {\bibfnamefont {Y.}~\bibnamefont
  {Ge}}, \bibinfo {author} {\bibfnamefont {W.}~\bibnamefont {Wan}}, \bibinfo
  {author} {\bibfnamefont {F.}~\bibnamefont {Yang}}, \ and\ \bibinfo {author}
  {\bibfnamefont {Y.}~\bibnamefont {Yao}},\ }\bibfield  {title} {\enquote
  {\bibinfo {title} {{The strain effect on superconductivity in phosphorene: a
  first-principles prediction}},}\ }\href {\doibase
  10.1088/1367-2630/17/3/035008} {\bibfield  {journal} {\bibinfo  {journal}
  {New J. Phys.}\ }\textbf {\bibinfo {volume} {17}},\ \bibinfo {pages} {035008}
  (\bibinfo {year} {2015})}\BibitemShut {NoStop}%
\bibitem [{\citenamefont {Chapman}\ \emph {et~al.}(2016)\citenamefont
  {Chapman}, \citenamefont {Su}, \citenamefont {Howard}, \citenamefont
  {Kundys}, \citenamefont {Grigorenko}, \citenamefont {Guinea}, \citenamefont
  {Geim}, \citenamefont {Grigorieva},\ and\ \citenamefont
  {Nair}}]{Chapman2016}%
  \BibitemOpen
  \bibfield  {author} {\bibinfo {author} {\bibfnamefont {J.}~\bibnamefont
  {Chapman}}, \bibinfo {author} {\bibfnamefont {Y.}~\bibnamefont {Su}},
  \bibinfo {author} {\bibfnamefont {C.~A.}\ \bibnamefont {Howard}}, \bibinfo
  {author} {\bibfnamefont {D.}~\bibnamefont {Kundys}}, \bibinfo {author}
  {\bibfnamefont {A.~N.}\ \bibnamefont {Grigorenko}}, \bibinfo {author}
  {\bibfnamefont {F.}~\bibnamefont {Guinea}}, \bibinfo {author} {\bibfnamefont
  {A.~K.}\ \bibnamefont {Geim}}, \bibinfo {author} {\bibfnamefont {I.~V.}\
  \bibnamefont {Grigorieva}}, \ and\ \bibinfo {author} {\bibfnamefont {R.~R.}\
  \bibnamefont {Nair}},\ }\bibfield  {title} {\enquote {\bibinfo {title}
  {{Superconductivity in Ca-doped graphene laminates}},}\ }\href {\doibase
  10.1038/srep23254} {\bibfield  {journal} {\bibinfo  {journal} {Sci. Rep.}\
  }\textbf {\bibinfo {volume} {6}},\ \bibinfo {pages} {23254} (\bibinfo {year}
  {2016})}\BibitemShut {NoStop}%
\bibitem [{\citenamefont {Zhang}\ \emph {et~al.}(2017)\citenamefont {Zhang},
  \citenamefont {Waters}, \citenamefont {Geim},\ and\ \citenamefont
  {Grigorieva}}]{Zhang2017}%
  \BibitemOpen
  \bibfield  {author} {\bibinfo {author} {\bibfnamefont {R.}~\bibnamefont
  {Zhang}}, \bibinfo {author} {\bibfnamefont {J.}~\bibnamefont {Waters}},
  \bibinfo {author} {\bibfnamefont {A.~K.}\ \bibnamefont {Geim}}, \ and\
  \bibinfo {author} {\bibfnamefont {I.~V.}\ \bibnamefont {Grigorieva}},\
  }\bibfield  {title} {\enquote {\bibinfo {title} {{Intercalant-independent
  transition temperature in superconducting black phosphorus}},}\ }\href
  {\doibase 10.1038/ncomms15036} {\bibfield  {journal} {\bibinfo  {journal}
  {Nat. Commun.}\ }\textbf {\bibinfo {volume} {8}},\ \bibinfo {pages} {15036}
  (\bibinfo {year} {2017})}\BibitemShut {NoStop}%
\bibitem [{\citenamefont {Lu}\ \emph {et~al.}(2018)\citenamefont {Lu},
  \citenamefont {Zheliuk}, \citenamefont {Chen}, \citenamefont {Leermakers},
  \citenamefont {Hussey}, \citenamefont {Zeitler},\ and\ \citenamefont
  {Ye}}]{Lu2018}%
  \BibitemOpen
  \bibfield  {author} {\bibinfo {author} {\bibfnamefont {J.}~\bibnamefont
  {Lu}}, \bibinfo {author} {\bibfnamefont {O.}~\bibnamefont {Zheliuk}},
  \bibinfo {author} {\bibfnamefont {Q.}~\bibnamefont {Chen}}, \bibinfo {author}
  {\bibfnamefont {I.}~\bibnamefont {Leermakers}}, \bibinfo {author}
  {\bibfnamefont {N.~E.}\ \bibnamefont {Hussey}}, \bibinfo {author}
  {\bibfnamefont {U.}~\bibnamefont {Zeitler}}, \ and\ \bibinfo {author}
  {\bibfnamefont {J.}~\bibnamefont {Ye}},\ }\bibfield  {title} {\enquote
  {\bibinfo {title} {{Full superconducting dome of strong Ising protection in
  gated monolayer WS$_2$.}}}\ }\href {\doibase 10.1073/pnas.1716781115}
  {\bibfield  {journal} {\bibinfo  {journal} {Proc. Natl. Acad. Sci. U. S. A.}\
  }\textbf {\bibinfo {volume} {115}},\ \bibinfo {pages} {3551} (\bibinfo {year}
  {2018})}\BibitemShut {NoStop}%
\bibitem [{\citenamefont {Xi}\ \emph {et~al.}(2015)\citenamefont {Xi},
  \citenamefont {Wang}, \citenamefont {Zhao}, \citenamefont {Park},
  \citenamefont {Law}, \citenamefont {Berger}, \citenamefont {Forr{\'{o}}},
  \citenamefont {Shan},\ and\ \citenamefont {Mak}}]{Xi2015}%
  \BibitemOpen
  \bibfield  {author} {\bibinfo {author} {\bibfnamefont {X.}~\bibnamefont
  {Xi}}, \bibinfo {author} {\bibfnamefont {Z.}~\bibnamefont {Wang}}, \bibinfo
  {author} {\bibfnamefont {W.}~\bibnamefont {Zhao}}, \bibinfo {author}
  {\bibfnamefont {J.-H.}\ \bibnamefont {Park}}, \bibinfo {author}
  {\bibfnamefont {K.~T.}\ \bibnamefont {Law}}, \bibinfo {author} {\bibfnamefont
  {H.}~\bibnamefont {Berger}}, \bibinfo {author} {\bibfnamefont
  {L.}~\bibnamefont {Forr{\'{o}}}}, \bibinfo {author} {\bibfnamefont
  {J.}~\bibnamefont {Shan}}, \ and\ \bibinfo {author} {\bibfnamefont {K.~F.}\
  \bibnamefont {Mak}},\ }\bibfield  {title} {\enquote {\bibinfo {title} {{Ising
  pairing in superconducting NbSe$_2$ atomic layers}},}\ }\href
  {http://dx.doi.org/10.1038/nphys3538 http://10.0.4.14/nphys3538} {\bibfield
  {journal} {\bibinfo  {journal} {Nat. Phys.}\ }\textbf {\bibinfo {volume}
  {12}},\ \bibinfo {pages} {139} (\bibinfo {year} {2015})}\BibitemShut
  {NoStop}%
\bibitem [{\citenamefont {Lu}\ \emph {et~al.}(2015)\citenamefont {Lu},
  \citenamefont {Zheliuk}, \citenamefont {Leermakers}, \citenamefont {Yuan},
  \citenamefont {Zeitler}, \citenamefont {Law},\ and\ \citenamefont
  {Ye}}]{Lu2015}%
  \BibitemOpen
  \bibfield  {author} {\bibinfo {author} {\bibfnamefont {J.~M.}\ \bibnamefont
  {Lu}}, \bibinfo {author} {\bibfnamefont {O.}~\bibnamefont {Zheliuk}},
  \bibinfo {author} {\bibfnamefont {I.}~\bibnamefont {Leermakers}}, \bibinfo
  {author} {\bibfnamefont {N.~F.~Q.}\ \bibnamefont {Yuan}}, \bibinfo {author}
  {\bibfnamefont {U.}~\bibnamefont {Zeitler}}, \bibinfo {author} {\bibfnamefont
  {K.~T.}\ \bibnamefont {Law}}, \ and\ \bibinfo {author} {\bibfnamefont
  {J.~T.}\ \bibnamefont {Ye}},\ }\bibfield  {title} {\enquote {\bibinfo {title}
  {Evidence for two-dimensional ising superconductivity in gated mos$_2$.}}\
  }\href {\doibase 10.1126/science.aab2277} {\bibfield  {journal} {\bibinfo
  {journal} {Science}\ }\textbf {\bibinfo {volume} {350}},\ \bibinfo {pages}
  {1353} (\bibinfo {year} {2015})}\BibitemShut {NoStop}%
\bibitem [{\citenamefont {Ye}\ \emph {et~al.}(2012)\citenamefont {Ye},
  \citenamefont {Zhang}, \citenamefont {Akashi}, \citenamefont {Bahramy},
  \citenamefont {Arita},\ and\ \citenamefont {Iwasa}}]{Ye1193}%
  \BibitemOpen
  \bibfield  {author} {\bibinfo {author} {\bibfnamefont {J.~T.}\ \bibnamefont
  {Ye}}, \bibinfo {author} {\bibfnamefont {Y.~J.}\ \bibnamefont {Zhang}},
  \bibinfo {author} {\bibfnamefont {R.}~\bibnamefont {Akashi}}, \bibinfo
  {author} {\bibfnamefont {M.~S.}\ \bibnamefont {Bahramy}}, \bibinfo {author}
  {\bibfnamefont {R.}~\bibnamefont {Arita}}, \ and\ \bibinfo {author}
  {\bibfnamefont {Y.}~\bibnamefont {Iwasa}},\ }\bibfield  {title} {\enquote
  {\bibinfo {title} {{Superconducting Dome in a Gate-Tuned Band Insulator}},}\
  }\href {\doibase 10.1126/science.1228006} {\bibfield  {journal} {\bibinfo
  {journal} {Science}\ }\textbf {\bibinfo {volume} {338}},\ \bibinfo {pages}
  {1193} (\bibinfo {year} {2012})}\BibitemShut {NoStop}%
\bibitem [{\citenamefont {Heersche}\ \emph {et~al.}()\citenamefont {Heersche},
  \citenamefont {Jarillo-Herrero}, \citenamefont {Oostinga}, \citenamefont
  {Vandersypen},\ and\ \citenamefont {Morpurgo}}]{Heersche2007}%
  \BibitemOpen
  \bibfield  {author} {\bibinfo {author} {\bibfnamefont {H.~B.}\ \bibnamefont
  {Heersche}}, \bibinfo {author} {\bibfnamefont {P.~J.}\ \bibnamefont
  {Jarillo-Herrero}}, \bibinfo {author} {\bibfnamefont {B.}~\bibnamefont
  {Oostinga}}, \bibinfo {author} {\bibfnamefont {L.~M.~K.}\ \bibnamefont
  {Vandersypen}}, \ and\ \bibinfo {author} {\bibfnamefont {A.~F.}\ \bibnamefont
  {Morpurgo}},\ }\href@noop {} {\ }\BibitemShut {NoStop}%
\bibitem [{\citenamefont {Yabuki}\ \emph {et~al.}(2016)\citenamefont {Yabuki},
  \citenamefont {Moriya}, \citenamefont {Arai}, \citenamefont {Sata},
  \citenamefont {Morikawa}, \citenamefont {Masubuchi},\ and\ \citenamefont
  {Machida}}]{Yabuki2016}%
  \BibitemOpen
  \bibfield  {author} {\bibinfo {author} {\bibfnamefont {N.}~\bibnamefont
  {Yabuki}}, \bibinfo {author} {\bibfnamefont {R.}~\bibnamefont {Moriya}},
  \bibinfo {author} {\bibfnamefont {M.}~\bibnamefont {Arai}}, \bibinfo {author}
  {\bibfnamefont {Y.}~\bibnamefont {Sata}}, \bibinfo {author} {\bibfnamefont
  {S.}~\bibnamefont {Morikawa}}, \bibinfo {author} {\bibfnamefont
  {S.}~\bibnamefont {Masubuchi}}, \ and\ \bibinfo {author} {\bibfnamefont
  {T.}~\bibnamefont {Machida}},\ }\bibfield  {title} {\enquote {\bibinfo
  {title} {{Supercurrent in van der Waals Josephson junction}},}\ }\href
  {\doibase 10.1038/ncomms10616} {\bibfield  {journal} {\bibinfo  {journal}
  {Nat. Commun.}\ }\textbf {\bibinfo {volume} {7}},\ \bibinfo {pages} {10616}
  (\bibinfo {year} {2016})}\BibitemShut {NoStop}%
\bibitem [{\citenamefont {Profeta}\ \emph {et~al.}(2012)\citenamefont
  {Profeta}, \citenamefont {Calandra},\ and\ \citenamefont
  {Mauri}}]{Profeta2012}%
  \BibitemOpen
  \bibfield  {author} {\bibinfo {author} {\bibfnamefont {G.}~\bibnamefont
  {Profeta}}, \bibinfo {author} {\bibfnamefont {M.}~\bibnamefont {Calandra}}, \
  and\ \bibinfo {author} {\bibfnamefont {F.}~\bibnamefont {Mauri}},\ }\bibfield
   {title} {\enquote {\bibinfo {title} {{Phonon-mediated superconductivity in
  graphene by lithium deposition}},}\ }\href {\doibase 10.1038/nphys2181}
  {\bibfield  {journal} {\bibinfo  {journal} {Nat. Phys.}\ }\textbf {\bibinfo
  {volume} {8}},\ \bibinfo {pages} {131} (\bibinfo {year} {2012})}\BibitemShut
  {NoStop}%
\bibitem [{\citenamefont {Margine}\ and\ \citenamefont
  {Giustino}(2014)}]{Margine2014}%
  \BibitemOpen
  \bibfield  {author} {\bibinfo {author} {\bibfnamefont {E.~R.}\ \bibnamefont
  {Margine}}\ and\ \bibinfo {author} {\bibfnamefont {F.}~\bibnamefont
  {Giustino}},\ }\bibfield  {title} {\enquote {\bibinfo {title} {{Two-gap
  superconductivity in heavily n-doped graphene: Ab initio Migdal-Eliashberg
  theory}},}\ }\href {\doibase 10.1103/PhysRevB.90.014518} {\bibfield
  {journal} {\bibinfo  {journal} {Phys. Rev. B}\ }\textbf {\bibinfo {volume}
  {90}},\ \bibinfo {pages} {014518} (\bibinfo {year} {2014})}\BibitemShut
  {NoStop}%
\bibitem [{\citenamefont {Shao}\ \emph {et~al.}(2014)\citenamefont {Shao},
  \citenamefont {Lu}, \citenamefont {Lv},\ and\ \citenamefont
  {Sun}}]{Shao2014}%
  \BibitemOpen
  \bibfield  {author} {\bibinfo {author} {\bibfnamefont {D.~F.}\ \bibnamefont
  {Shao}}, \bibinfo {author} {\bibfnamefont {W.~J.}\ \bibnamefont {Lu}},
  \bibinfo {author} {\bibfnamefont {H.~Y.}\ \bibnamefont {Lv}}, \ and\ \bibinfo
  {author} {\bibfnamefont {Y.~P.}\ \bibnamefont {Sun}},\ }\bibfield  {title}
  {\enquote {\bibinfo {title} {{Electron-doped phosphorene: A potential
  monolayer superconductor}},}\ }\href {\doibase 10.1209/0295-5075/108/67004}
  {\bibfield  {journal} {\bibinfo  {journal} {EPL}\ }\textbf {\bibinfo {volume}
  {108}},\ \bibinfo {pages} {67004} (\bibinfo {year} {2014})}\BibitemShut
  {NoStop}%
\bibitem [{\citenamefont {Kong}\ \emph {et~al.}(2018)\citenamefont {Kong},
  \citenamefont {Gao}, \citenamefont {Yan}, \citenamefont {Lu},\ and\
  \citenamefont {Xiang}}]{Kong2018}%
  \BibitemOpen
  \bibfield  {author} {\bibinfo {author} {\bibfnamefont {X.}~\bibnamefont
  {Kong}}, \bibinfo {author} {\bibfnamefont {M.}~\bibnamefont {Gao}}, \bibinfo
  {author} {\bibfnamefont {X.-W.}\ \bibnamefont {Yan}}, \bibinfo {author}
  {\bibfnamefont {Z.-Y.}\ \bibnamefont {Lu}}, \ and\ \bibinfo {author}
  {\bibfnamefont {T.}~\bibnamefont {Xiang}},\ }\bibfield  {title} {\enquote
  {\bibinfo {title} {{Superconductivity in electron-doped arsenene}},}\ }\href
  {https://arxiv.org/abs/1801.00545} {\  (\bibinfo {year} {2018})},\ \Eprint
  {http://arxiv.org/abs/1801.00545} {arXiv:1801.00545} \BibitemShut {NoStop}%
\bibitem [{\citenamefont {Durajski}\ \emph {et~al.}(2014)\citenamefont
  {Durajski}, \citenamefont {Szczȩ{\'{s}}niak},\ and\ \citenamefont
  {Szczȩ{\'{s}}niak}}]{Durajski2014}%
  \BibitemOpen
  \bibfield  {author} {\bibinfo {author} {\bibfnamefont {A.~P.}\ \bibnamefont
  {Durajski}}, \bibinfo {author} {\bibfnamefont {D.}~\bibnamefont
  {Szczȩ{\'{s}}niak}}, \ and\ \bibinfo {author} {\bibfnamefont
  {R.}~\bibnamefont {Szczȩ{\'{s}}niak}},\ }\bibfield  {title} {\enquote
  {\bibinfo {title} {{Study of the superconducting phase in silicene under
  biaxial tensile strain}},}\ }\href {\doibase 10.1016/J.SSC.2014.09.007}
  {\bibfield  {journal} {\bibinfo  {journal} {Solid State Commun.}\ }\textbf
  {\bibinfo {volume} {200}},\ \bibinfo {pages} {17} (\bibinfo {year}
  {2014})}\BibitemShut {NoStop}%
\bibitem [{\citenamefont {Ge}\ and\ \citenamefont {Liu}(2013)}]{Ge2013}%
  \BibitemOpen
  \bibfield  {author} {\bibinfo {author} {\bibfnamefont {Y.}~\bibnamefont
  {Ge}}\ and\ \bibinfo {author} {\bibfnamefont {A.~Y.}\ \bibnamefont {Liu}},\
  }\bibfield  {title} {\enquote {\bibinfo {title} {{Phonon-mediated
  superconductivity in electron-doped single-layer MoS$_{2}$: A
  first-principles prediction}},}\ }\href {\doibase 10.1103/PhysRevB.87.241408}
  {\bibfield  {journal} {\bibinfo  {journal} {Phys. Rev. B}\ }\textbf {\bibinfo
  {volume} {87}},\ \bibinfo {pages} {241408} (\bibinfo {year}
  {2013})}\BibitemShut {NoStop}%
\bibitem [{\citenamefont {Hohenberg}\ and\ \citenamefont
  {Kohn}(1964)}]{Hohenberg1964}%
  \BibitemOpen
  \bibfield  {author} {\bibinfo {author} {\bibfnamefont {P.}~\bibnamefont
  {Hohenberg}}\ and\ \bibinfo {author} {\bibfnamefont {W.}~\bibnamefont
  {Kohn}},\ }\bibfield  {title} {\enquote {\bibinfo {title} {{Inhomogeneous
  Electron Gas}},}\ }\href {\doibase 10.1103/PhysRev.136.B864} {\bibfield
  {journal} {\bibinfo  {journal} {Phys. Rev.}\ }\textbf {\bibinfo {volume}
  {136}},\ \bibinfo {pages} {B864} (\bibinfo {year} {1964})}\BibitemShut
  {NoStop}%
\bibitem [{\citenamefont {Kohn}\ and\ \citenamefont {Sham}(1965)}]{Kohn1965}%
  \BibitemOpen
  \bibfield  {author} {\bibinfo {author} {\bibfnamefont {W.}~\bibnamefont
  {Kohn}}\ and\ \bibinfo {author} {\bibfnamefont {L.~J.}\ \bibnamefont
  {Sham}},\ }\bibfield  {title} {\enquote {\bibinfo {title} {{Self-Consistent
  Equations Including Exchange and Correlation Effects}},}\ }\href {\doibase
  10.1103/PhysRev.140.A1133} {\bibfield  {journal} {\bibinfo  {journal} {Phys.
  Rev.}\ }\textbf {\bibinfo {volume} {140}},\ \bibinfo {pages} {A1133}
  (\bibinfo {year} {1965})}\BibitemShut {NoStop}%
\bibitem [{\citenamefont {Gonze}\ and\ \citenamefont {Lee}(1997)}]{Gonze1997}%
  \BibitemOpen
  \bibfield  {author} {\bibinfo {author} {\bibfnamefont {X.}~\bibnamefont
  {Gonze}}\ and\ \bibinfo {author} {\bibfnamefont {C.}~\bibnamefont {Lee}},\
  }\bibfield  {title} {\enquote {\bibinfo {title} {{Dynamical matrices, Born
  effective charges, dielectric permittivity tensors, and interatomic force
  constants from density-functional perturbation theory}},}\ }\href {\doibase
  10.1103/PhysRevB.55.10355} {\bibfield  {journal} {\bibinfo  {journal} {Phys.
  Rev. B}\ }\textbf {\bibinfo {volume} {55}},\ \bibinfo {pages} {10355}
  (\bibinfo {year} {1997})}\BibitemShut {NoStop}%
\bibitem [{\citenamefont {Giustino}\ \emph {et~al.}(2007)\citenamefont
  {Giustino}, \citenamefont {Cohen},\ and\ \citenamefont
  {Louie}}]{Giustino2007}%
  \BibitemOpen
  \bibfield  {author} {\bibinfo {author} {\bibfnamefont {F.}~\bibnamefont
  {Giustino}}, \bibinfo {author} {\bibfnamefont {M.~L.}\ \bibnamefont {Cohen}},
  \ and\ \bibinfo {author} {\bibfnamefont {S.~G.}\ \bibnamefont {Louie}},\
  }\bibfield  {title} {\enquote {\bibinfo {title} {{Electron-phonon interaction
  using Wannier functions}},}\ }\href {\doibase 10.1103/PhysRevB.76.165108}
  {\bibfield  {journal} {\bibinfo  {journal} {Phys. Rev. B}\ }\textbf {\bibinfo
  {volume} {76}},\ \bibinfo {pages} {165108} (\bibinfo {year}
  {2007})}\BibitemShut {NoStop}%
\bibitem [{\citenamefont {Marzari}\ \emph {et~al.}(2012)\citenamefont
  {Marzari}, \citenamefont {Mostofi}, \citenamefont {Yates}, \citenamefont
  {Souza},\ and\ \citenamefont {Vanderbilt}}]{Marzari2012}%
  \BibitemOpen
  \bibfield  {author} {\bibinfo {author} {\bibfnamefont {N.}~\bibnamefont
  {Marzari}}, \bibinfo {author} {\bibfnamefont {A.~A.}\ \bibnamefont
  {Mostofi}}, \bibinfo {author} {\bibfnamefont {J.~R.}\ \bibnamefont {Yates}},
  \bibinfo {author} {\bibfnamefont {I.}~\bibnamefont {Souza}}, \ and\ \bibinfo
  {author} {\bibfnamefont {D.}~\bibnamefont {Vanderbilt}},\ }\bibfield  {title}
  {\enquote {\bibinfo {title} {{Maximally localized Wannier functions: Theory
  and applications}},}\ }\href {\doibase 10.1103/RevModPhys.84.1419} {\bibfield
   {journal} {\bibinfo  {journal} {Rev. Mod. Phys.}\ }\textbf {\bibinfo
  {volume} {84}},\ \bibinfo {pages} {1419} (\bibinfo {year}
  {2012})}\BibitemShut {NoStop}%
\bibitem [{\citenamefont {Ni}\ \emph {et~al.}(2012)\citenamefont {Ni},
  \citenamefont {Liu}, \citenamefont {Tang}, \citenamefont {Zheng},
  \citenamefont {Zhou}, \citenamefont {Qin}, \citenamefont {Gao}, \citenamefont
  {Yu},\ and\ \citenamefont {Lu}}]{Ni2012}%
  \BibitemOpen
  \bibfield  {author} {\bibinfo {author} {\bibfnamefont {Z.}~\bibnamefont
  {Ni}}, \bibinfo {author} {\bibfnamefont {Q.}~\bibnamefont {Liu}}, \bibinfo
  {author} {\bibfnamefont {K.}~\bibnamefont {Tang}}, \bibinfo {author}
  {\bibfnamefont {J.}~\bibnamefont {Zheng}}, \bibinfo {author} {\bibfnamefont
  {J.}~\bibnamefont {Zhou}}, \bibinfo {author} {\bibfnamefont {R.}~\bibnamefont
  {Qin}}, \bibinfo {author} {\bibfnamefont {Z.}~\bibnamefont {Gao}}, \bibinfo
  {author} {\bibfnamefont {D.}~\bibnamefont {Yu}}, \ and\ \bibinfo {author}
  {\bibfnamefont {J.}~\bibnamefont {Lu}},\ }\bibfield  {title} {\enquote
  {\bibinfo {title} {{Tunable Bandgap in Silicene and Germanene}},}\ }\href
  {\doibase 10.1021/nl203065e} {\bibfield  {journal} {\bibinfo  {journal} {Nano
  Lett.}\ }\textbf {\bibinfo {volume} {12}},\ \bibinfo {pages} {113} (\bibinfo
  {year} {2012})}\BibitemShut {NoStop}%
\bibitem [{\citenamefont {Acun}\ \emph {et~al.}(2015)\citenamefont {Acun},
  \citenamefont {Zhang}, \citenamefont {Bampoulis}, \citenamefont {Farmanbar},
  \citenamefont {van Houselt}, \citenamefont {Rudenko}, \citenamefont
  {Lingenfelder}, \citenamefont {Brocks}, \citenamefont {Poelsema},
  \citenamefont {Katsnelson},\ and\ \citenamefont {Zandvliet}}]{Acun2015}%
  \BibitemOpen
  \bibfield  {author} {\bibinfo {author} {\bibfnamefont {A.}~\bibnamefont
  {Acun}}, \bibinfo {author} {\bibfnamefont {L.}~\bibnamefont {Zhang}},
  \bibinfo {author} {\bibfnamefont {P.}~\bibnamefont {Bampoulis}}, \bibinfo
  {author} {\bibfnamefont {M.}~\bibnamefont {Farmanbar}}, \bibinfo {author}
  {\bibfnamefont {A.}~\bibnamefont {van Houselt}}, \bibinfo {author}
  {\bibfnamefont {A.~N.}\ \bibnamefont {Rudenko}}, \bibinfo {author}
  {\bibfnamefont {M.}~\bibnamefont {Lingenfelder}}, \bibinfo {author}
  {\bibfnamefont {G.}~\bibnamefont {Brocks}}, \bibinfo {author} {\bibfnamefont
  {B.}~\bibnamefont {Poelsema}}, \bibinfo {author} {\bibfnamefont {M.~I.}\
  \bibnamefont {Katsnelson}}, \ and\ \bibinfo {author} {\bibfnamefont
  {H.~J.~W.}\ \bibnamefont {Zandvliet}},\ }\bibfield  {title} {\enquote
  {\bibinfo {title} {{Germanene: the germanium analogue of graphene}},}\ }\href
  {\doibase 10.1088/0953-8984/27/44/443002} {\bibfield  {journal} {\bibinfo
  {journal} {J. Phys. Condens. Matter}\ }\textbf {\bibinfo {volume} {27}},\
  \bibinfo {pages} {443002} (\bibinfo {year} {2015})}\BibitemShut {NoStop}%
\bibitem [{\citenamefont {Rudenko}\ \emph {et~al.}(2015)\citenamefont
  {Rudenko}, \citenamefont {Yuan},\ and\ \citenamefont
  {Katsnelson}}]{Rudenko2015}%
  \BibitemOpen
  \bibfield  {author} {\bibinfo {author} {\bibfnamefont {A.~N.}\ \bibnamefont
  {Rudenko}}, \bibinfo {author} {\bibfnamefont {S.}~\bibnamefont {Yuan}}, \
  and\ \bibinfo {author} {\bibfnamefont {M.~I.}\ \bibnamefont {Katsnelson}},\
  }\bibfield  {title} {\enquote {\bibinfo {title} {{Toward a realistic
  description of multilayer black phosphorus: From $GW$ approximation to
  large-scale tight-binding simulations}},}\ }\href {\doibase
  10.1103/PhysRevB.92.085419} {\bibfield  {journal} {\bibinfo  {journal} {Phys.
  Rev. B}\ }\textbf {\bibinfo {volume} {92}},\ \bibinfo {pages} {085419}
  (\bibinfo {year} {2015})}\BibitemShut {NoStop}%
\bibitem [{\citenamefont {Kim}\ \emph {et~al.}(2015)\citenamefont {Kim},
  \citenamefont {Baik}, \citenamefont {Ryu}, \citenamefont {Sohn},
  \citenamefont {Park}, \citenamefont {Park}, \citenamefont {Denlinger},
  \citenamefont {Yi}, \citenamefont {Choi},\ and\ \citenamefont
  {Kim}}]{Kim723}%
  \BibitemOpen
  \bibfield  {author} {\bibinfo {author} {\bibfnamefont {J.}~\bibnamefont
  {Kim}}, \bibinfo {author} {\bibfnamefont {S.~S.}\ \bibnamefont {Baik}},
  \bibinfo {author} {\bibfnamefont {S.~H.}\ \bibnamefont {Ryu}}, \bibinfo
  {author} {\bibfnamefont {Y.}~\bibnamefont {Sohn}}, \bibinfo {author}
  {\bibfnamefont {S.}~\bibnamefont {Park}}, \bibinfo {author} {\bibfnamefont
  {B.-G.}\ \bibnamefont {Park}}, \bibinfo {author} {\bibfnamefont
  {J.}~\bibnamefont {Denlinger}}, \bibinfo {author} {\bibfnamefont
  {Y.}~\bibnamefont {Yi}}, \bibinfo {author} {\bibfnamefont {H.~J.}\
  \bibnamefont {Choi}}, \ and\ \bibinfo {author} {\bibfnamefont {K.~S.}\
  \bibnamefont {Kim}},\ }\bibfield  {title} {\enquote {\bibinfo {title}
  {Observation of tunable band gap and anisotropic dirac semimetal state in
  black phosphorus},}\ }\href {\doibase 10.1126/science.aaa6486} {\bibfield
  {journal} {\bibinfo  {journal} {Science}\ }\textbf {\bibinfo {volume}
  {349}},\ \bibinfo {pages} {723} (\bibinfo {year} {2015})}\BibitemShut
  {NoStop}%
\bibitem [{\citenamefont {Liu}\ \emph {et~al.}(2015)\citenamefont {Liu},
  \citenamefont {Zhang}, \citenamefont {Abdalla}, \citenamefont {Fazzio},\ and\
  \citenamefont {Zunger}}]{Liu2015}%
  \BibitemOpen
  \bibfield  {author} {\bibinfo {author} {\bibfnamefont {Q.}~\bibnamefont
  {Liu}}, \bibinfo {author} {\bibfnamefont {X.}~\bibnamefont {Zhang}}, \bibinfo
  {author} {\bibfnamefont {L.~B.}\ \bibnamefont {Abdalla}}, \bibinfo {author}
  {\bibfnamefont {A.}~\bibnamefont {Fazzio}}, \ and\ \bibinfo {author}
  {\bibfnamefont {A.}~\bibnamefont {Zunger}},\ }\bibfield  {title} {\enquote
  {\bibinfo {title} {{Switching a Normal Insulator into a Topological Insulator
  via Electric Field with Application to Phosphorene}},}\ }\href {\doibase
  10.1021/nl5043769} {\bibfield  {journal} {\bibinfo  {journal} {Nano Lett.}\
  }\textbf {\bibinfo {volume} {15}},\ \bibinfo {pages} {1222} (\bibinfo {year}
  {2015})}\BibitemShut {NoStop}%
\bibitem [{\citenamefont {Korm{\'{a}}nyos}\ \emph {et~al.}(2013)\citenamefont
  {Korm{\'{a}}nyos}, \citenamefont {Z{\'{o}}lyomi}, \citenamefont {Drummond},
  \citenamefont {Rakyta}, \citenamefont {Burkard},\ and\ \citenamefont
  {Fal'ko}}]{Kormanyos2013}%
  \BibitemOpen
  \bibfield  {author} {\bibinfo {author} {\bibfnamefont {A.}~\bibnamefont
  {Korm{\'{a}}nyos}}, \bibinfo {author} {\bibfnamefont {V.}~\bibnamefont
  {Z{\'{o}}lyomi}}, \bibinfo {author} {\bibfnamefont {N.~D.}\ \bibnamefont
  {Drummond}}, \bibinfo {author} {\bibfnamefont {P.}~\bibnamefont {Rakyta}},
  \bibinfo {author} {\bibfnamefont {G.}~\bibnamefont {Burkard}}, \ and\
  \bibinfo {author} {\bibfnamefont {V.~I.}\ \bibnamefont {Fal'ko}},\ }\bibfield
   {title} {\enquote {\bibinfo {title} {{Monolayer MoS$_{2}$ : Trigonal
  warping, the $\Gamma$ valley, and spin-orbit coupling effects}},}\ }\href
  {\doibase 10.1103/PhysRevB.88.045416} {\bibfield  {journal} {\bibinfo
  {journal} {Phys. Rev. B}\ }\textbf {\bibinfo {volume} {88}},\ \bibinfo
  {pages} {045416} (\bibinfo {year} {2013})}\BibitemShut {NoStop}%
\bibitem [{\citenamefont {Giustino}(2017)}]{Giustino2017}%
  \BibitemOpen
  \bibfield  {author} {\bibinfo {author} {\bibfnamefont {F.}~\bibnamefont
  {Giustino}},\ }\bibfield  {title} {\enquote {\bibinfo {title}
  {{Electron-phonon interactions from first principles}},}\ }\href {\doibase
  10.1103/RevModPhys.89.015003} {\bibfield  {journal} {\bibinfo  {journal}
  {Rev. Mod. Phys.}\ }\textbf {\bibinfo {volume} {89}},\ \bibinfo {pages}
  {015003} (\bibinfo {year} {2017})}\BibitemShut {NoStop}%
\bibitem [{\citenamefont {Bazhirov}\ \emph {et~al.}(2010)\citenamefont
  {Bazhirov}, \citenamefont {Noffsinger},\ and\ \citenamefont
  {Cohen}}]{Bazhirov2010}%
  \BibitemOpen
  \bibfield  {author} {\bibinfo {author} {\bibfnamefont {T.}~\bibnamefont
  {Bazhirov}}, \bibinfo {author} {\bibfnamefont {J.}~\bibnamefont
  {Noffsinger}}, \ and\ \bibinfo {author} {\bibfnamefont {M.~L.}\ \bibnamefont
  {Cohen}},\ }\bibfield  {title} {\enquote {\bibinfo {title}
  {{Superconductivity and electron-phonon coupling in lithium at high
  pressures}},}\ }\href {\doibase 10.1103/PhysRevB.82.184509} {\bibfield
  {journal} {\bibinfo  {journal} {Phys. Rev. B}\ }\textbf {\bibinfo {volume}
  {82}},\ \bibinfo {pages} {184509} (\bibinfo {year} {2010})}\BibitemShut
  {NoStop}%
\bibitem [{\citenamefont {Katsnelson}\ \emph {et~al.}(1994)\citenamefont
  {Katsnelson}, \citenamefont {Naumov},\ and\ \citenamefont
  {Trefilov}}]{Katsnelson1994}%
  \BibitemOpen
  \bibfield  {author} {\bibinfo {author} {\bibfnamefont {M.~I.}\ \bibnamefont
  {Katsnelson}}, \bibinfo {author} {\bibfnamefont {I.~I.}\ \bibnamefont
  {Naumov}}, \ and\ \bibinfo {author} {\bibfnamefont {A.~V.}\ \bibnamefont
  {Trefilov}},\ }\bibfield  {title} {\enquote {\bibinfo {title} {{Singularities
  of the electronic structure and pre-martensitic anomalies of lattice
  properties in $\beta$-phases of metals and alloys}},}\ }\href {\doibase
  10.1080/01411599408201172} {\bibfield  {journal} {\bibinfo  {journal} {Phase
  Transit.}\ }\textbf {\bibinfo {volume} {49}},\ \bibinfo {pages} {143}
  (\bibinfo {year} {1994})}\BibitemShut {NoStop}%
\bibitem [{\citenamefont {McMillan}(1968)}]{McMillan1968}%
  \BibitemOpen
  \bibfield  {author} {\bibinfo {author} {\bibfnamefont {W.~L.}\ \bibnamefont
  {McMillan}},\ }\bibfield  {title} {\enquote {\bibinfo {title} {{Transition
  Temperature of Strong-Coupled Superconductors}},}\ }\href {\doibase
  10.1103/PhysRev.167.331} {\bibfield  {journal} {\bibinfo  {journal} {Phys.
  Rev.}\ }\textbf {\bibinfo {volume} {167}},\ \bibinfo {pages} {331} (\bibinfo
  {year} {1968})}\BibitemShut {NoStop}%
\bibitem [{\citenamefont {Allen}\ and\ \citenamefont
  {Dynes}(1975)}]{Allen1975}%
  \BibitemOpen
  \bibfield  {author} {\bibinfo {author} {\bibfnamefont {P.~B.}\ \bibnamefont
  {Allen}}\ and\ \bibinfo {author} {\bibfnamefont {R.~C.}\ \bibnamefont
  {Dynes}},\ }\bibfield  {title} {\enquote {\bibinfo {title} {{Transition
  temperature of strong-coupled superconductors reanalyzed}},}\ }\href
  {\doibase 10.1103/PhysRevB.12.905} {\bibfield  {journal} {\bibinfo  {journal}
  {Phys. Rev. B}\ }\textbf {\bibinfo {volume} {12}},\ \bibinfo {pages} {905}
  (\bibinfo {year} {1975})}\BibitemShut {NoStop}%
\bibitem [{\citenamefont {Migdal}(1958)}]{Migdal1958}%
  \BibitemOpen
  \bibfield  {author} {\bibinfo {author} {\bibfnamefont {A.~B.}\ \bibnamefont
  {Migdal}},\ }\bibfield  {title} {\enquote {\bibinfo {title} {{Interactions
  between electrons and lattice vibrations in a normal metal}},}\ }\href
  {http://www.jetp.ac.ru/cgi-bin/dn/e{\_}007{\_}06{\_}0996.pdf} {\bibfield
  {journal} {\bibinfo  {journal} {Sov. Phys. JETP}\ }\textbf {\bibinfo {volume}
  {34}},\ \bibinfo {pages} {1438} (\bibinfo {year} {1958})}\BibitemShut
  {NoStop}%
\bibitem [{\citenamefont {Eliashberg}(1960)}]{Eliashberg1960}%
  \BibitemOpen
  \bibfield  {author} {\bibinfo {author} {\bibfnamefont {G.~M.}\ \bibnamefont
  {Eliashberg}},\ }\bibfield  {title} {\enquote {\bibinfo {title}
  {{Interactions between electrons and lattice vibrations in a
  superconductor}},}\ }\href
  {http://www.jetp.ac.ru/cgi-bin/dn/e{\_}011{\_}03{\_}0696.pdf} {\bibfield
  {journal} {\bibinfo  {journal} {Sov. Phys. JETP}\ }\textbf {\bibinfo {volume}
  {11}},\ \bibinfo {pages} {966} (\bibinfo {year} {1960})}\BibitemShut
  {NoStop}%
\bibitem [{\citenamefont {Morel}\ and\ \citenamefont
  {Anderson}(1962)}]{Morel1962}%
  \BibitemOpen
  \bibfield  {author} {\bibinfo {author} {\bibfnamefont {P.}~\bibnamefont
  {Morel}}\ and\ \bibinfo {author} {\bibfnamefont {P.~W.}\ \bibnamefont
  {Anderson}},\ }\bibfield  {title} {\enquote {\bibinfo {title} {{Calculation
  of the Superconducting State Parameters with Retarded Electron-Phonon
  Interaction}},}\ }\href {\doibase 10.1103/PhysRev.125.1263} {\bibfield
  {journal} {\bibinfo  {journal} {Phys. Rev.}\ }\textbf {\bibinfo {volume}
  {125}},\ \bibinfo {pages} {1263} (\bibinfo {year} {1962})}\BibitemShut
  {NoStop}%
\bibitem [{\citenamefont {Peters}\ \emph {et~al.}(2018)\citenamefont {Peters},
  \citenamefont {Rudenko},\ and\ \citenamefont {Katsnelson}}]{Peters2018}%
  \BibitemOpen
  \bibfield  {author} {\bibinfo {author} {\bibfnamefont {L.}~\bibnamefont
  {Peters}}, \bibinfo {author} {\bibfnamefont {A.~N.}\ \bibnamefont {Rudenko}},
  \ and\ \bibinfo {author} {\bibfnamefont {M.~I.}\ \bibnamefont {Katsnelson}},\
  }\bibfield  {title} {\enquote {\bibinfo {title} {{\textit{Ab initio }study of
  the electron-phonon coupling at the Cr(001) surface}},}\ }\href {\doibase
  10.1103/PhysRevB.97.165438} {\bibfield  {journal} {\bibinfo  {journal} {Phys.
  Rev. B}\ }\textbf {\bibinfo {volume} {97}},\ \bibinfo {pages} {165438}
  (\bibinfo {year} {2018})}\BibitemShut {NoStop}%
\bibitem [{\citenamefont {Margine}\ \emph {et~al.}(2016)\citenamefont
  {Margine}, \citenamefont {Lambert},\ and\ \citenamefont
  {Giustino}}]{Margine2016}%
  \BibitemOpen
  \bibfield  {author} {\bibinfo {author} {\bibfnamefont {E.~R.}\ \bibnamefont
  {Margine}}, \bibinfo {author} {\bibfnamefont {H.}~\bibnamefont {Lambert}}, \
  and\ \bibinfo {author} {\bibfnamefont {F.}~\bibnamefont {Giustino}},\
  }\bibfield  {title} {\enquote {\bibinfo {title} {{Electron-phonon interaction
  and pairing mechanism in superconducting Ca-intercalated bilayer
  graphene}},}\ }\href {\doibase 10.1038/srep21414} {\bibfield  {journal}
  {\bibinfo  {journal} {Scientific Reports}\ }\textbf {\bibinfo {volume} {6}},\
  \bibinfo {pages} {21414} (\bibinfo {year} {2016})}\BibitemShut {NoStop}%
\bibitem [{\citenamefont {Heil}\ \emph {et~al.}(2017)\citenamefont {Heil},
  \citenamefont {Ponc\'e}, \citenamefont {Lambert}, \citenamefont {Schlipf},
  \citenamefont {Margine},\ and\ \citenamefont {Giustino}}]{Heil2017}%
  \BibitemOpen
  \bibfield  {author} {\bibinfo {author} {\bibfnamefont {C.}~\bibnamefont
  {Heil}}, \bibinfo {author} {\bibfnamefont {S.}~\bibnamefont {Ponc\'e}},
  \bibinfo {author} {\bibfnamefont {H.}~\bibnamefont {Lambert}}, \bibinfo
  {author} {\bibfnamefont {M.}~\bibnamefont {Schlipf}}, \bibinfo {author}
  {\bibfnamefont {E.~R.}\ \bibnamefont {Margine}}, \ and\ \bibinfo {author}
  {\bibfnamefont {F.}~\bibnamefont {Giustino}},\ }\bibfield  {title} {\enquote
  {\bibinfo {title} {Origin of superconductivity and latent charge density wave
  in ${\mathrm{nbs}}_{2}$},}\ }\href {\doibase 10.1103/PhysRevLett.119.087003}
  {\bibfield  {journal} {\bibinfo  {journal} {Phys. Rev. Lett.}\ }\textbf
  {\bibinfo {volume} {119}},\ \bibinfo {pages} {087003} (\bibinfo {year}
  {2017})}\BibitemShut {NoStop}%
\bibitem [{\citenamefont {Giannozzi}\ \emph {et~al.}(2009)\citenamefont
  {Giannozzi}, \citenamefont {Baroni}, \citenamefont {Bonini}, \citenamefont
  {Calandra}, \citenamefont {Car}, \citenamefont {Cavazzoni}, \citenamefont
  {Ceresoli}, \citenamefont {Chiarotti}, \citenamefont {Cococcioni},
  \citenamefont {Dabo}, \citenamefont {Corso}, \citenamefont {Fabris},
  \citenamefont {Fratesi}, \citenamefont {de~Gironcoli}, \citenamefont
  {Gebauer}, \citenamefont {Gerstmann}, \citenamefont {Gougoussis},
  \citenamefont {Kokalj}, \citenamefont {Lazzeri}, \citenamefont
  {Martin-Samos}, \citenamefont {Marzari}, \citenamefont {Mauri}, \citenamefont
  {Mazzarello}, \citenamefont {Paolini}, \citenamefont {Pasquarello},
  \citenamefont {Paulatto}, \citenamefont {Sbraccia}, \citenamefont {Scandolo},
  \citenamefont {Sclauzero}, \citenamefont {Seitsonen}, \citenamefont
  {Smogunov}, \citenamefont {Umari},\ and\ \citenamefont
  {Wentzcovitch}}]{Giannozzi2009}%
  \BibitemOpen
  \bibfield  {author} {\bibinfo {author} {\bibfnamefont {P.}~\bibnamefont
  {Giannozzi}}, \bibinfo {author} {\bibfnamefont {S.}~\bibnamefont {Baroni}},
  \bibinfo {author} {\bibfnamefont {N.}~\bibnamefont {Bonini}}, \bibinfo
  {author} {\bibfnamefont {M.}~\bibnamefont {Calandra}}, \bibinfo {author}
  {\bibfnamefont {R.}~\bibnamefont {Car}}, \bibinfo {author} {\bibfnamefont
  {C.}~\bibnamefont {Cavazzoni}}, \bibinfo {author} {\bibfnamefont
  {D.}~\bibnamefont {Ceresoli}}, \bibinfo {author} {\bibfnamefont {G.~L.}\
  \bibnamefont {Chiarotti}}, \bibinfo {author} {\bibfnamefont {M.}~\bibnamefont
  {Cococcioni}}, \bibinfo {author} {\bibfnamefont {I.}~\bibnamefont {Dabo}},
  \bibinfo {author} {\bibfnamefont {A.~D.}\ \bibnamefont {Corso}}, \bibinfo
  {author} {\bibfnamefont {S.}~\bibnamefont {Fabris}}, \bibinfo {author}
  {\bibfnamefont {G.}~\bibnamefont {Fratesi}}, \bibinfo {author} {\bibfnamefont
  {S.}~\bibnamefont {de~Gironcoli}}, \bibinfo {author} {\bibfnamefont
  {R.}~\bibnamefont {Gebauer}}, \bibinfo {author} {\bibfnamefont
  {U.}~\bibnamefont {Gerstmann}}, \bibinfo {author} {\bibfnamefont
  {C.}~\bibnamefont {Gougoussis}}, \bibinfo {author} {\bibfnamefont
  {A.}~\bibnamefont {Kokalj}}, \bibinfo {author} {\bibfnamefont
  {M.}~\bibnamefont {Lazzeri}}, \bibinfo {author} {\bibfnamefont
  {L.}~\bibnamefont {Martin-Samos}}, \bibinfo {author} {\bibfnamefont
  {N.}~\bibnamefont {Marzari}}, \bibinfo {author} {\bibfnamefont
  {F.}~\bibnamefont {Mauri}}, \bibinfo {author} {\bibfnamefont
  {R.}~\bibnamefont {Mazzarello}}, \bibinfo {author} {\bibfnamefont
  {S.}~\bibnamefont {Paolini}}, \bibinfo {author} {\bibfnamefont
  {A.}~\bibnamefont {Pasquarello}}, \bibinfo {author} {\bibfnamefont
  {L.}~\bibnamefont {Paulatto}}, \bibinfo {author} {\bibfnamefont
  {C.}~\bibnamefont {Sbraccia}}, \bibinfo {author} {\bibfnamefont
  {S.}~\bibnamefont {Scandolo}}, \bibinfo {author} {\bibfnamefont
  {G.}~\bibnamefont {Sclauzero}}, \bibinfo {author} {\bibfnamefont {A.~P.}\
  \bibnamefont {Seitsonen}}, \bibinfo {author} {\bibfnamefont {A.}~\bibnamefont
  {Smogunov}}, \bibinfo {author} {\bibfnamefont {P.}~\bibnamefont {Umari}}, \
  and\ \bibinfo {author} {\bibfnamefont {R.~M.}\ \bibnamefont {Wentzcovitch}},\
  }\bibfield  {title} {\enquote {\bibinfo {title} {{QUANTUM ESPRESSO: a modular
  and open-source software project for quantum simulations of materials}},}\
  }\href {http://stacks.iop.org/0953-8984/21/i=39/a=395502} {\bibfield
  {journal} {\bibinfo  {journal} {J. Phys. Condens. Matter}\ }\textbf {\bibinfo
  {volume} {21}},\ \bibinfo {pages} {395502} (\bibinfo {year}
  {2009})}\BibitemShut {NoStop}%
\bibitem [{\citenamefont {Giannozzi}\ \emph {et~al.}(2017)\citenamefont
  {Giannozzi}, \citenamefont {Andreussi}, \citenamefont {Brumme}, \citenamefont
  {Bunau}, \citenamefont {{Buongiorno Nardelli}}, \citenamefont {Calandra},
  \citenamefont {Car}, \citenamefont {Cavazzoni}, \citenamefont {Ceresoli},
  \citenamefont {Cococcioni}, \citenamefont {Colonna}, \citenamefont
  {Carnimeo}, \citenamefont {{Dal Corso}}, \citenamefont {de~Gironcoli},
  \citenamefont {Delugas}, \citenamefont {DiStasio}, \citenamefont {Ferretti},
  \citenamefont {Floris}, \citenamefont {Fratesi}, \citenamefont {Fugallo},
  \citenamefont {Gebauer}, \citenamefont {Gerstmann}, \citenamefont {Giustino},
  \citenamefont {Gorni}, \citenamefont {Jia}, \citenamefont {Kawamura},
  \citenamefont {Ko}, \citenamefont {Kokalj}, \citenamefont
  {K{\"{u}}{\c{c}}{\"{u}}kbenli}, \citenamefont {Lazzeri}, \citenamefont
  {Marsili}, \citenamefont {Marzari}, \citenamefont {Mauri}, \citenamefont
  {Nguyen}, \citenamefont {Nguyen}, \citenamefont {Otero-de-la Roza},
  \citenamefont {Paulatto}, \citenamefont {Ponc{\'{e}}}, \citenamefont {Rocca},
  \citenamefont {Sabatini}, \citenamefont {Santra}, \citenamefont {Schlipf},
  \citenamefont {Seitsonen}, \citenamefont {Smogunov}, \citenamefont {Timrov},
  \citenamefont {Thonhauser}, \citenamefont {Umari}, \citenamefont {Vast},
  \citenamefont {Wu},\ and\ \citenamefont {Baroni}}]{Giannozzi2017}%
  \BibitemOpen
  \bibfield  {author} {\bibinfo {author} {\bibfnamefont {P.}~\bibnamefont
  {Giannozzi}}, \bibinfo {author} {\bibfnamefont {O.}~\bibnamefont
  {Andreussi}}, \bibinfo {author} {\bibfnamefont {T.}~\bibnamefont {Brumme}},
  \bibinfo {author} {\bibfnamefont {O.}~\bibnamefont {Bunau}}, \bibinfo
  {author} {\bibfnamefont {M.}~\bibnamefont {{Buongiorno Nardelli}}}, \bibinfo
  {author} {\bibfnamefont {M.}~\bibnamefont {Calandra}}, \bibinfo {author}
  {\bibfnamefont {R.}~\bibnamefont {Car}}, \bibinfo {author} {\bibfnamefont
  {C.}~\bibnamefont {Cavazzoni}}, \bibinfo {author} {\bibfnamefont
  {D.}~\bibnamefont {Ceresoli}}, \bibinfo {author} {\bibfnamefont
  {M.}~\bibnamefont {Cococcioni}}, \bibinfo {author} {\bibfnamefont
  {N.}~\bibnamefont {Colonna}}, \bibinfo {author} {\bibfnamefont
  {I.}~\bibnamefont {Carnimeo}}, \bibinfo {author} {\bibfnamefont
  {A.}~\bibnamefont {{Dal Corso}}}, \bibinfo {author} {\bibfnamefont
  {S.}~\bibnamefont {de~Gironcoli}}, \bibinfo {author} {\bibfnamefont
  {P.}~\bibnamefont {Delugas}}, \bibinfo {author} {\bibfnamefont {R.~A.}\
  \bibnamefont {DiStasio}}, \bibinfo {author} {\bibfnamefont {A.}~\bibnamefont
  {Ferretti}}, \bibinfo {author} {\bibfnamefont {A.}~\bibnamefont {Floris}},
  \bibinfo {author} {\bibfnamefont {G.}~\bibnamefont {Fratesi}}, \bibinfo
  {author} {\bibfnamefont {G.}~\bibnamefont {Fugallo}}, \bibinfo {author}
  {\bibfnamefont {R.}~\bibnamefont {Gebauer}}, \bibinfo {author} {\bibfnamefont
  {U.}~\bibnamefont {Gerstmann}}, \bibinfo {author} {\bibfnamefont
  {F.}~\bibnamefont {Giustino}}, \bibinfo {author} {\bibfnamefont
  {T.}~\bibnamefont {Gorni}}, \bibinfo {author} {\bibfnamefont
  {J.}~\bibnamefont {Jia}}, \bibinfo {author} {\bibfnamefont {M.}~\bibnamefont
  {Kawamura}}, \bibinfo {author} {\bibfnamefont {H.-Y.}\ \bibnamefont {Ko}},
  \bibinfo {author} {\bibfnamefont {A.}~\bibnamefont {Kokalj}}, \bibinfo
  {author} {\bibfnamefont {E.}~\bibnamefont {K{\"{u}}{\c{c}}{\"{u}}kbenli}},
  \bibinfo {author} {\bibfnamefont {M.}~\bibnamefont {Lazzeri}}, \bibinfo
  {author} {\bibfnamefont {M.}~\bibnamefont {Marsili}}, \bibinfo {author}
  {\bibfnamefont {N.}~\bibnamefont {Marzari}}, \bibinfo {author} {\bibfnamefont
  {F.}~\bibnamefont {Mauri}}, \bibinfo {author} {\bibfnamefont {N.~L.}\
  \bibnamefont {Nguyen}}, \bibinfo {author} {\bibfnamefont {H.-V.}\
  \bibnamefont {Nguyen}}, \bibinfo {author} {\bibfnamefont {A.}~\bibnamefont
  {Otero-de-la Roza}}, \bibinfo {author} {\bibfnamefont {L.}~\bibnamefont
  {Paulatto}}, \bibinfo {author} {\bibfnamefont {S.}~\bibnamefont
  {Ponc{\'{e}}}}, \bibinfo {author} {\bibfnamefont {D.}~\bibnamefont {Rocca}},
  \bibinfo {author} {\bibfnamefont {R.}~\bibnamefont {Sabatini}}, \bibinfo
  {author} {\bibfnamefont {B.}~\bibnamefont {Santra}}, \bibinfo {author}
  {\bibfnamefont {M.}~\bibnamefont {Schlipf}}, \bibinfo {author} {\bibfnamefont
  {A.~P.}\ \bibnamefont {Seitsonen}}, \bibinfo {author} {\bibfnamefont
  {A.}~\bibnamefont {Smogunov}}, \bibinfo {author} {\bibfnamefont
  {I.}~\bibnamefont {Timrov}}, \bibinfo {author} {\bibfnamefont
  {T.}~\bibnamefont {Thonhauser}}, \bibinfo {author} {\bibfnamefont
  {P.}~\bibnamefont {Umari}}, \bibinfo {author} {\bibfnamefont
  {N.}~\bibnamefont {Vast}}, \bibinfo {author} {\bibfnamefont {X.}~\bibnamefont
  {Wu}}, \ and\ \bibinfo {author} {\bibfnamefont {S.}~\bibnamefont {Baroni}},\
  }\bibfield  {title} {\enquote {\bibinfo {title} {{Advanced capabilities for
  materials modelling with Quantum ESPRESSO}},}\ }\href {\doibase
  10.1088/1361-648X/aa8f79} {\bibfield  {journal} {\bibinfo  {journal} {J.
  Phys. Condens. Matter}\ }\textbf {\bibinfo {volume} {29}},\ \bibinfo {pages}
  {465901} (\bibinfo {year} {2017})}\BibitemShut {NoStop}%
\bibitem [{\citenamefont {Monkhorst}\ and\ \citenamefont
  {Pack}(1976)}]{Monkhorst1976}%
  \BibitemOpen
  \bibfield  {author} {\bibinfo {author} {\bibfnamefont {H.~J.}\ \bibnamefont
  {Monkhorst}}\ and\ \bibinfo {author} {\bibfnamefont {J.~D.}\ \bibnamefont
  {Pack}},\ }\bibfield  {title} {\enquote {\bibinfo {title} {{Special points
  for Brillouin-zone integrations}},}\ }\href {\doibase
  10.1103/PhysRevB.13.5188} {\bibfield  {journal} {\bibinfo  {journal} {Phys.
  Rev. B}\ }\textbf {\bibinfo {volume} {13}},\ \bibinfo {pages} {5188}
  (\bibinfo {year} {1976})}\BibitemShut {NoStop}%
\bibitem [{\citenamefont {Ponc{\'{e}}}\ \emph {et~al.}(2016)\citenamefont
  {Ponc{\'{e}}}, \citenamefont {Margine}, \citenamefont {Verdi},\ and\
  \citenamefont {Giustino}}]{Ponce2016}%
  \BibitemOpen
  \bibfield  {author} {\bibinfo {author} {\bibfnamefont {S.}~\bibnamefont
  {Ponc{\'{e}}}}, \bibinfo {author} {\bibfnamefont {E.~R.}\ \bibnamefont
  {Margine}}, \bibinfo {author} {\bibfnamefont {C.}~\bibnamefont {Verdi}}, \
  and\ \bibinfo {author} {\bibfnamefont {F.}~\bibnamefont {Giustino}},\
  }\bibfield  {title} {\enquote {\bibinfo {title} {{EPW: Electron–phonon
  coupling, transport and superconducting properties using maximally localized
  Wannier functions}},}\ }\href {\doibase 10.1016/J.CPC.2016.07.028} {\bibfield
   {journal} {\bibinfo  {journal} {Comput. Phys. Commun.}\ }\textbf {\bibinfo
  {volume} {209}},\ \bibinfo {pages} {116} (\bibinfo {year}
  {2016})}\BibitemShut {NoStop}%
\bibitem [{\citenamefont {Sohier}\ \emph {et~al.}(2017)\citenamefont {Sohier},
  \citenamefont {Calandra},\ and\ \citenamefont {Mauri}}]{Sohier2017}%
  \BibitemOpen
  \bibfield  {author} {\bibinfo {author} {\bibfnamefont {T.}~\bibnamefont
  {Sohier}}, \bibinfo {author} {\bibfnamefont {M.}~\bibnamefont {Calandra}}, \
  and\ \bibinfo {author} {\bibfnamefont {F.}~\bibnamefont {Mauri}},\ }\bibfield
   {title} {\enquote {\bibinfo {title} {{Density functional perturbation theory
  for gated two-dimensional heterostructures: Theoretical developments and
  application to flexural phonons in graphene}},}\ }\href {\doibase
  10.1103/PhysRevB.96.075448} {\bibfield  {journal} {\bibinfo  {journal} {Phys.
  Rev. B}\ }\textbf {\bibinfo {volume} {96}},\ \bibinfo {pages} {075448}
  (\bibinfo {year} {2017})}\BibitemShut {NoStop}%
\bibitem [{\citenamefont {Daghero}\ \emph {et~al.}(2012)\citenamefont
  {Daghero}, \citenamefont {Paolucci}, \citenamefont {Sola}, \citenamefont
  {Tortello}, \citenamefont {Ummarino}, \citenamefont {Agosto}, \citenamefont
  {Gonnelli}, \citenamefont {Nair},\ and\ \citenamefont
  {Gerbaldi}}]{Daghero2012}%
  \BibitemOpen
  \bibfield  {author} {\bibinfo {author} {\bibfnamefont {D.}~\bibnamefont
  {Daghero}}, \bibinfo {author} {\bibfnamefont {F.}~\bibnamefont {Paolucci}},
  \bibinfo {author} {\bibfnamefont {A.}~\bibnamefont {Sola}}, \bibinfo {author}
  {\bibfnamefont {M.}~\bibnamefont {Tortello}}, \bibinfo {author}
  {\bibfnamefont {G.~A.}\ \bibnamefont {Ummarino}}, \bibinfo {author}
  {\bibfnamefont {M.}~\bibnamefont {Agosto}}, \bibinfo {author} {\bibfnamefont
  {R.~S.}\ \bibnamefont {Gonnelli}}, \bibinfo {author} {\bibfnamefont {J.~R.}\
  \bibnamefont {Nair}}, \ and\ \bibinfo {author} {\bibfnamefont
  {C.}~\bibnamefont {Gerbaldi}},\ }\bibfield  {title} {\enquote {\bibinfo
  {title} {Large conductance modulation of gold thin films by huge charge
  injection via electrochemical gating},}\ }\href {\doibase
  10.1103/PhysRevLett.108.066807} {\bibfield  {journal} {\bibinfo  {journal}
  {Phys. Rev. Lett.}\ }\textbf {\bibinfo {volume} {108}},\ \bibinfo {pages}
  {066807} (\bibinfo {year} {2012})}\BibitemShut {NoStop}%
\bibitem [{\citenamefont {Efetov}\ and\ \citenamefont
  {Kim}(2010)}]{Efetov2010}%
  \BibitemOpen
  \bibfield  {author} {\bibinfo {author} {\bibfnamefont {D.~K.}\ \bibnamefont
  {Efetov}}\ and\ \bibinfo {author} {\bibfnamefont {P.}~\bibnamefont {Kim}},\
  }\bibfield  {title} {\enquote {\bibinfo {title} {{Controlling Electron-Phonon
  Interactions in Graphene at Ultrahigh Carrier Densities}},}\ }\href {\doibase
  10.1103/PhysRevLett.105.256805} {\bibfield  {journal} {\bibinfo  {journal}
  {Phys. Rev. Lett.}\ }\textbf {\bibinfo {volume} {105}},\ \bibinfo {pages}
  {256805} (\bibinfo {year} {2010})}\BibitemShut {NoStop}%
\bibitem [{\citenamefont {Xu}\ \emph {et~al.}(2017)\citenamefont {Xu},
  \citenamefont {Lu}, \citenamefont {Kinder}, \citenamefont {Seabaugh},\ and\
  \citenamefont {Fullerton-Shirey}}]{Xu2017}%
  \BibitemOpen
  \bibfield  {author} {\bibinfo {author} {\bibfnamefont {K.}~\bibnamefont
  {Xu}}, \bibinfo {author} {\bibfnamefont {H.}~\bibnamefont {Lu}}, \bibinfo
  {author} {\bibfnamefont {E.~W.}\ \bibnamefont {Kinder}}, \bibinfo {author}
  {\bibfnamefont {A.}~\bibnamefont {Seabaugh}}, \ and\ \bibinfo {author}
  {\bibfnamefont {S.~K.}\ \bibnamefont {Fullerton-Shirey}},\ }\bibfield
  {title} {\enquote {\bibinfo {title} {{Monolayer Solid-State Electrolyte for
  Electric Double Layer Gating of Graphene Field-Effect Transistors}},}\ }\href
  {\doibase 10.1021/acsnano.6b08505} {\bibfield  {journal} {\bibinfo  {journal}
  {ACS Nano}\ }\textbf {\bibinfo {volume} {11}},\ \bibinfo {pages} {5453}
  (\bibinfo {year} {2017})}\BibitemShut {NoStop}%
\bibitem [{\citenamefont {Antropov}\ \emph {et~al.}(1988)\citenamefont
  {Antropov}, \citenamefont {Vaks}, \citenamefont {Katsnel'son}, \citenamefont
  {Koreshkov}, \citenamefont {Likhtenshteĭn},\ and\ \citenamefont
  {Trefilov}}]{Antropov1988}%
  \BibitemOpen
  \bibfield  {author} {\bibinfo {author} {\bibfnamefont {V.~I.}\ \bibnamefont
  {Antropov}}, \bibinfo {author} {\bibfnamefont {V.~G.}\ \bibnamefont {Vaks}},
  \bibinfo {author} {\bibfnamefont {M.~I.}\ \bibnamefont {Katsnel'son}},
  \bibinfo {author} {\bibfnamefont {V.~G.}\ \bibnamefont {Koreshkov}}, \bibinfo
  {author} {\bibfnamefont {A.~I.}\ \bibnamefont {Likhtenshteĭn}}, \ and\
  \bibinfo {author} {\bibfnamefont {A.~V.}\ \bibnamefont {Trefilov}},\
  }\bibfield  {title} {\enquote {\bibinfo {title} {{Effect of proximity of the
  Fermi level to singular points in the band structure on the kinetic and
  lattice properties of metals and alloys}},}\ }\href {\doibase
  10.1070/PU1988v031n03ABEH005725} {\bibfield  {journal} {\bibinfo  {journal}
  {Sov. Phys. Uspekhi}\ }\textbf {\bibinfo {volume} {31}},\ \bibinfo {pages}
  {278} (\bibinfo {year} {1988})}\BibitemShut {NoStop}%
\bibitem [{\citenamefont {Vaks}\ \emph {et~al.}(1989)\citenamefont {Vaks},
  \citenamefont {Katsnelson}, \citenamefont {Koreshkov}, \citenamefont
  {Likhtenstein}, \citenamefont {Parfenov}, \citenamefont {Skok}, \citenamefont
  {Sukhoparov}, \citenamefont {Trefilov},\ and\ \citenamefont
  {Chernyshov}}]{Vaks1989}%
  \BibitemOpen
  \bibfield  {author} {\bibinfo {author} {\bibfnamefont {V.~G.}\ \bibnamefont
  {Vaks}}, \bibinfo {author} {\bibfnamefont {M.~I.}\ \bibnamefont
  {Katsnelson}}, \bibinfo {author} {\bibfnamefont {V.~G.}\ \bibnamefont
  {Koreshkov}}, \bibinfo {author} {\bibfnamefont {A.~I.}\ \bibnamefont
  {Likhtenstein}}, \bibinfo {author} {\bibfnamefont {O.~E.}\ \bibnamefont
  {Parfenov}}, \bibinfo {author} {\bibfnamefont {V.~F.}\ \bibnamefont {Skok}},
  \bibinfo {author} {\bibfnamefont {V.~A.}\ \bibnamefont {Sukhoparov}},
  \bibinfo {author} {\bibfnamefont {A.~V.}\ \bibnamefont {Trefilov}}, \ and\
  \bibinfo {author} {\bibfnamefont {A.~A.}\ \bibnamefont {Chernyshov}},\
  }\bibfield  {title} {\enquote {\bibinfo {title} {{An experimental and
  theoretical study of martensitic phase transitions in Li and Na under
  pressure}},}\ }\href {\doibase 10.1088/0953-8984/1/32/001} {\bibfield
  {journal} {\bibinfo  {journal} {J. Phys. Condens. Matter}\ }\textbf {\bibinfo
  {volume} {1}},\ \bibinfo {pages} {5319} (\bibinfo {year} {1989})}\BibitemShut
  {NoStop}%
\bibitem [{\citenamefont {Li}\ \emph {et~al.}(2018)\citenamefont {Li},
  \citenamefont {Partoens},\ and\ \citenamefont {Peeters}}]{Li2018}%
  \BibitemOpen
  \bibfield  {author} {\bibinfo {author} {\bibfnamefont {L.~L.}\ \bibnamefont
  {Li}}, \bibinfo {author} {\bibfnamefont {B.}~\bibnamefont {Partoens}}, \ and\
  \bibinfo {author} {\bibfnamefont {F.~M.}\ \bibnamefont {Peeters}},\
  }\bibfield  {title} {\enquote {\bibinfo {title} {Tuning the electronic
  properties of gated multilayer phosphorene: A self-consistent tight-binding
  study},}\ }\href {\doibase 10.1103/PhysRevB.97.155424} {\bibfield  {journal}
  {\bibinfo  {journal} {Phys. Rev. B}\ }\textbf {\bibinfo {volume} {97}},\
  \bibinfo {pages} {155424} (\bibinfo {year} {2018})}\BibitemShut {NoStop}%
\bibitem [{\citenamefont {Wang}\ \emph {et~al.}(2015)\citenamefont {Wang},
  \citenamefont {Pandey},\ and\ \citenamefont {Karna}}]{Wang2015}%
  \BibitemOpen
  \bibfield  {author} {\bibinfo {author} {\bibfnamefont {G.}~\bibnamefont
  {Wang}}, \bibinfo {author} {\bibfnamefont {R.}~\bibnamefont {Pandey}}, \ and\
  \bibinfo {author} {\bibfnamefont {S.~P.}\ \bibnamefont {Karna}},\ }\bibfield
  {title} {\enquote {\bibinfo {title} {{Atomically thin group V elemental
  films: Theoretical investigations of antimonene allotropes}},}\ }\href
  {\doibase 10.1021/acsami.5b02441} {\bibfield  {journal} {\bibinfo  {journal}
  {ACS Appl. Mater. Interfaces}\ }\textbf {\bibinfo {volume} {7}},\ \bibinfo
  {pages} {11490} (\bibinfo {year} {2015})}\BibitemShut {NoStop}%
\bibitem [{\citenamefont {Landa}\ \emph {et~al.}(2018)\citenamefont {Landa},
  \citenamefont {S{\"{o}}derlind}, \citenamefont {Naumov}, \citenamefont
  {Klepeis},\ and\ \citenamefont {Vitos}}]{Landa2018}%
  \BibitemOpen
  \bibfield  {author} {\bibinfo {author} {\bibfnamefont {A.}~\bibnamefont
  {Landa}}, \bibinfo {author} {\bibfnamefont {P.}~\bibnamefont
  {S{\"{o}}derlind}}, \bibinfo {author} {\bibfnamefont {I.}~\bibnamefont
  {Naumov}}, \bibinfo {author} {\bibfnamefont {J.}~\bibnamefont {Klepeis}}, \
  and\ \bibinfo {author} {\bibfnamefont {L.}~\bibnamefont {Vitos}},\ }\bibfield
   {title} {\enquote {\bibinfo {title} {{Kohn Anomaly and Phase Stability in
  Group VB Transition Metals}},}\ }\href {\doibase 10.3390/computation6020029}
  {\bibfield  {journal} {\bibinfo  {journal} {Computation}\ }\textbf {\bibinfo
  {volume} {6}},\ \bibinfo {pages} {29} (\bibinfo {year} {2018})}\BibitemShut
  {NoStop}%
\bibitem [{\citenamefont {Wang}\ \emph {et~al.}(2016)\citenamefont {Wang},
  \citenamefont {Wang},\ and\ \citenamefont {Zhao}}]{SWang2016}%
  \BibitemOpen
  \bibfield  {author} {\bibinfo {author} {\bibfnamefont {S.}~\bibnamefont
  {Wang}}, \bibinfo {author} {\bibfnamefont {W.}~\bibnamefont {Wang}}, \ and\
  \bibinfo {author} {\bibfnamefont {G.}~\bibnamefont {Zhao}},\ }\bibfield
  {title} {\enquote {\bibinfo {title} {{Thermal transport properties of
  antimonene: an ab initio study}},}\ }\href {\doibase 10.1039/C6CP06088A}
  {\bibfield  {journal} {\bibinfo  {journal} {Phys. Chem. Chem. Phys.}\
  }\textbf {\bibinfo {volume} {18}},\ \bibinfo {pages} {31217} (\bibinfo {year}
  {2016})}\BibitemShut {NoStop}%
\bibitem [{\citenamefont {Liu}\ \emph {et~al.}(2016)\citenamefont {Liu},
  \citenamefont {Every},\ and\ \citenamefont {Tom{\'{a}}nek}}]{DLiu2016}%
  \BibitemOpen
  \bibfield  {author} {\bibinfo {author} {\bibfnamefont {D.}~\bibnamefont
  {Liu}}, \bibinfo {author} {\bibfnamefont {A.~G.}\ \bibnamefont {Every}}, \
  and\ \bibinfo {author} {\bibfnamefont {D.}~\bibnamefont {Tom{\'{a}}nek}},\
  }\bibfield  {title} {\enquote {\bibinfo {title} {{Continuum approach for
  long-wavelength acoustic phonons in quasi-two-dimensional structures}},}\
  }\href {\doibase 10.1103/PhysRevB.94.165432} {\bibfield  {journal} {\bibinfo
  {journal} {Phys. Rev. B}\ }\textbf {\bibinfo {volume} {94}},\ \bibinfo
  {pages} {165432} (\bibinfo {year} {2016})}\BibitemShut {NoStop}%
\bibitem [{\citenamefont {Zhu}\ and\ \citenamefont
  {Tom\'anek}(2014)}]{Zhu2014}%
  \BibitemOpen
  \bibfield  {author} {\bibinfo {author} {\bibfnamefont {Z.}~\bibnamefont
  {Zhu}}\ and\ \bibinfo {author} {\bibfnamefont {D.}~\bibnamefont
  {Tom\'anek}},\ }\bibfield  {title} {\enquote {\bibinfo {title}
  {Semiconducting layered blue phosphorus: A computational study},}\ }\href
  {\doibase 10.1103/PhysRevLett.112.176802} {\bibfield  {journal} {\bibinfo
  {journal} {Phys. Rev. Lett.}\ }\textbf {\bibinfo {volume} {112}},\ \bibinfo
  {pages} {176802} (\bibinfo {year} {2014})}\BibitemShut {NoStop}%
\bibitem [{\citenamefont {Rudenko}\ \emph {et~al.}(2016)\citenamefont
  {Rudenko}, \citenamefont {Brener},\ and\ \citenamefont
  {Katsnelson}}]{Rudenko2016}%
  \BibitemOpen
  \bibfield  {author} {\bibinfo {author} {\bibfnamefont {A.~N.}\ \bibnamefont
  {Rudenko}}, \bibinfo {author} {\bibfnamefont {S.}~\bibnamefont {Brener}}, \
  and\ \bibinfo {author} {\bibfnamefont {M.~I.}\ \bibnamefont {Katsnelson}},\
  }\bibfield  {title} {\enquote {\bibinfo {title} {{Intrinsic Charge Carrier
  Mobility in Single-Layer Black Phosphorus}},}\ }\href {\doibase
  10.1103/PhysRevLett.116.246401} {\bibfield  {journal} {\bibinfo  {journal}
  {Phys. Rev. Lett.}\ }\textbf {\bibinfo {volume} {116}},\ \bibinfo {pages}
  {246401} (\bibinfo {year} {2016})}\BibitemShut {NoStop}%
\bibitem [{\citenamefont {Katsnelson}(2010)}]{Katsnelson2010}%
  \BibitemOpen
  \bibfield  {author} {\bibinfo {author} {\bibfnamefont {M.~I.}\ \bibnamefont
  {Katsnelson}},\ }\bibfield  {title} {\enquote {\bibinfo {title} {{Flexuron: A
  self-trapped state of electron in crystalline membranes}},}\ }\href {\doibase
  10.1103/PhysRevB.82.205433} {\bibfield  {journal} {\bibinfo  {journal} {Phys.
  Rev. B}\ }\textbf {\bibinfo {volume} {82}},\ \bibinfo {pages} {205433}
  (\bibinfo {year} {2010})}\BibitemShut {NoStop}%
\bibitem [{\citenamefont {Zhao}\ \emph {et~al.}(2017)\citenamefont {Zhao},
  \citenamefont {Wang}, \citenamefont {Zhang}, \citenamefont {Lv},
  \citenamefont {Wu}, \citenamefont {Qiao}, \citenamefont {Song},\ and\
  \citenamefont {Gao}}]{Zhao2017}%
  \BibitemOpen
  \bibfield  {author} {\bibinfo {author} {\bibfnamefont {J.}~\bibnamefont
  {Zhao}}, \bibinfo {author} {\bibfnamefont {M.}~\bibnamefont {Wang}}, \bibinfo
  {author} {\bibfnamefont {X.}~\bibnamefont {Zhang}}, \bibinfo {author}
  {\bibfnamefont {Y.}~\bibnamefont {Lv}}, \bibinfo {author} {\bibfnamefont
  {T.}~\bibnamefont {Wu}}, \bibinfo {author} {\bibfnamefont {S.}~\bibnamefont
  {Qiao}}, \bibinfo {author} {\bibfnamefont {S.}~\bibnamefont {Song}}, \ and\
  \bibinfo {author} {\bibfnamefont {B.}~\bibnamefont {Gao}},\ }\bibfield
  {title} {\enquote {\bibinfo {title} {{Application of sodium-ion-based solid
  electrolyte in electrostatic tuning of carrier density in graphene}},}\
  }\href {\doibase 10.1038/s41598-017-03413-5} {\bibfield  {journal} {\bibinfo
  {journal} {Sci. Rep.}\ }\textbf {\bibinfo {volume} {7}},\ \bibinfo {pages}
  {3168} (\bibinfo {year} {2017})}\BibitemShut {NoStop}%
\bibitem [{\citenamefont {Kresin}(1987)}]{Kresin1987}%
  \BibitemOpen
  \bibfield  {author} {\bibinfo {author} {\bibfnamefont {V.~Z.}\ \bibnamefont
  {Kresin}},\ }\bibfield  {title} {\enquote {\bibinfo {title} {{On the critical
  temperature for any strength of the electron-phonon coupling}},}\ }\href
  {\doibase 10.1016/0375-9601(87)90744-4} {\bibfield  {journal} {\bibinfo
  {journal} {Phys. Lett. A}\ }\textbf {\bibinfo {volume} {122}},\ \bibinfo
  {pages} {434} (\bibinfo {year} {1987})}\BibitemShut {NoStop}%
\bibitem [{\citenamefont {Prishchenko}\ \emph {et~al.}(2018)\citenamefont
  {Prishchenko}, \citenamefont {Mazurenko}, \citenamefont {Katsnelson},\ and\
  \citenamefont {Rudenko}}]{Prishchenko2018}%
  \BibitemOpen
  \bibfield  {author} {\bibinfo {author} {\bibfnamefont {D.~A.}\ \bibnamefont
  {Prishchenko}}, \bibinfo {author} {\bibfnamefont {V.~G.}\ \bibnamefont
  {Mazurenko}}, \bibinfo {author} {\bibfnamefont {M.~I.}\ \bibnamefont
  {Katsnelson}}, \ and\ \bibinfo {author} {\bibfnamefont {A.~N.}\ \bibnamefont
  {Rudenko}},\ }\bibfield  {title} {\enquote {\bibinfo {title} {Gate-tunable
  infrared plasmons in electron-doped single-layer antimony},}\ }\href
  {\doibase 10.1103/PhysRevB.98.201401} {\bibfield  {journal} {\bibinfo
  {journal} {Phys. Rev. B}\ }\textbf {\bibinfo {volume} {98}},\ \bibinfo
  {pages} {201401} (\bibinfo {year} {2018})}\BibitemShut {NoStop}%
\bibitem [{\citenamefont {Zeng}\ \emph {et~al.}(2016)\citenamefont {Zeng},
  \citenamefont {Zhao}, \citenamefont {Li},\ and\ \citenamefont
  {Ni}}]{Zeng2016}%
  \BibitemOpen
  \bibfield  {author} {\bibinfo {author} {\bibfnamefont {S.}~\bibnamefont
  {Zeng}}, \bibinfo {author} {\bibfnamefont {Y.}~\bibnamefont {Zhao}}, \bibinfo
  {author} {\bibfnamefont {G.}~\bibnamefont {Li}}, \ and\ \bibinfo {author}
  {\bibfnamefont {J.}~\bibnamefont {Ni}},\ }\bibfield  {title} {\enquote
  {\bibinfo {title} {{Strongly enhanced superconductivity in doped monolayer
  MoS$_{2}$ by strain}},}\ }\href {\doibase 10.1103/PhysRevB.94.024501}
  {\bibfield  {journal} {\bibinfo  {journal} {Phys. Rev. B}\ }\textbf {\bibinfo
  {volume} {94}},\ \bibinfo {pages} {024501} (\bibinfo {year}
  {2016})}\BibitemShut {NoStop}%
\end{thebibliography}%

\end{document}